\documentclass[aps,prb,twocolumn,superscriptaddress]{revtex4}
\usepackage{graphicx}
\usepackage{latexsym}
\usepackage{amssymb}
\usepackage{amsmath}
\usepackage{amsfonts}
\usepackage{upgreek}
\usepackage{bm}
\usepackage{multirow}
\usepackage{enumitem}
\usepackage{color}
\usepackage{hyperref}
\hypersetup{
colorlinks = true,
linkcolor = [rgb]{0.70,0.13,0.13},
citecolor = [rgb]{0.13,0.55,0.13},
urlcolor = [rgb]{0.25, 0.41, 0.88}}
\newcommand{\bra}[1]{\langle #1|}
\newcommand{\ket}[1]{|#1 \rangle}

\newcommand{\ii}{\mathrm{i}}

\newcommand{\dsZ}{\mathbb{Z}}

\newcommand{\dsX}{\mathbb{X}}
\newcommand{\id}{\mathfrak{1}}
\newcommand{\cyc}{\mathfrak{c}}

\newcommand{\scE}{\mathcal{E}}
\newcommand{\scH}{\mathcal{H}}

\newcommand{\scO}{\mathcal{O}}

\newcommand{\scR}{\mathcal{R}}
\newcommand{\OTOC}{\mathsf{OTOC}}
\newcommand{\Wg}{\mathsf{Wg}}
\newcommand{\Tr}{\operatorname{Tr}}

\newcommand{\dia}[3]{\raisebox{#3pt}{\includegraphics[height=#2pt]{dia_#1}}}
\newcommand{\eq}[1]{\begin{equation}#1\end{equation}}
\newcommand{\eqs}[1]{\begin{equation}\begin{split}#1\end{split}\end{equation}}
\newcommand{\eqnref}[1]{Eq.\,\eqref{#1}}
\newcommand{\figref}[1]{Fig.\,\ref{#1}}
\newcommand{\tabref}[1]{Tab.\,\ref{#1}}

\newenvironment{defn}[1][Definition]{\begin{trivlist}
\item[\hskip \labelsep {\bfseries #1}]}{\end{trivlist}}

\begin{document}
\title{Entanglement Features of Random Hamiltonian Dynamics}
\author{Yi-Zhuang You}
\affiliation{Department of Physics, Harvard University, Cambridge, MA 02138, USA}
\affiliation{Department of Physics, University of California, San Diego, CA 92093, USA}
\author{Yingfei Gu}
\affiliation{Department of Physics, Harvard University, Cambridge, MA 02138, USA}
\date{\today}
\begin{abstract}

We introduce the concept of entanglement features of unitary gates, as a collection of exponentiated entanglement entropies over all bipartitions of input and output channels. We obtained the general formula for time-dependent $n$th-R\'enyi entanglement features for unitary gates generated by random Hamiltonian. In particular, we propose an Ising formulation for the 2nd-R\'enyi entanglement features of random Hamiltonian dynamics, which admits a holographic tensor network interpretation. As a general description of entanglement properties, we show that the entanglement features can be applied to several dynamical measures of thermalization, including the out-of-time-order correlation and the entanglement growth after a quantum quench. We also analyze the Yoshida-Kitaev probabilistic protocol for random Hamiltonian dynamics.

\end{abstract}
\maketitle
%\tableofcontents

\section{Introduction}

The dynamics of quantum many-body entanglement lies in the core of understanding thermalization, information scrambling and quantum chaos in many-body systems\cite{sekino2008fast, 
kim2013ballistic, 
liu2014entanglement, 
Kitaev2014hidden,
asplund2015entanglement, 
kaufman2016quantum,
maldacena2016bound, 
casini2016spread,
hosur2016chaos, 
aleiner2016microscopic,
gu2016fractional,
ho2017entanglement,
mezei2017entanglement,
mezei2017entanglement2,
gu2017spread,
leviatan2017quantum,
white2018quantum,
xu2018accessing}. 
Recently, there has been rapid progresses in the study of entanglement production and propagation in random unitary dynamics\cite{chandran2015finite,
nahum2017quantum,
nahum2017operator,
nahum2017dynamics,
von2017operator,
rakovszky2017diffusive,
khemani2017operator, 
jonay2018coarse}, where the time-evolution of quantum many-body systems is modeled by a unitary circuit in which all local unitary gates are independently random. The randomness in the unitary circuit efficiently removes the basis specific details and allows us to focus on the universal properties of entanglement dynamics. The same  philosophy also underlies the recent works\cite{hayden2016holographic,
qi2017holographic,
han2017discrete,
leutheusser2017tensor,
qi2018space,
you2018machine,
cotler2017superdensity,
vasseur2018entanglement}
 of using random tensor networks to model entangled many-body states or chaotic unitary evolutions. 
Due to the lack of time-translation symmetry in the random unitary circuit, energy is not conserved under \emph{random unitary dynamics}, which obscure its application to problems like energy transport. One step toward a generic quantum dynamics with energy conservation is to consider the \emph{random Hamiltonian dynamics}\cite{vznidarivc2012subsystem,cotler2017chaos,vijay2018finite}, i.e., unitary evolutions $U(t)=e^{-\ii H t}$ generated by time-independent random Hamiltonians $H$. 

In this work, we will consider the system of $N$ qudits. Each qudit corresponds to a $d$-dimensional local Hilbert space. The many-body Hilbert space is a direct product of qudit Hilbert spaces. The quantum dynamics of qudits is described by a random Hamiltonian that simultaneously acts on all qudits without locality. Although the Hamiltonian is non-local, the tensor product structure of the many-body Hilbert space still allows us to specify entanglement regions and to define the entanglement entropy over different partitions of qudits. 
%The non-locality of the Hamiltonian leads to fast scrambling of quantum information and strong thermalization. 
The goal of this work is to study the entanglement dynamics under the time-evolution generated by such non-local random Hamiltonians. 
Similar discussions of subsystem entanglement with non-local Hamiltonians also appear in the study of the Sachdev-Ye-Kitaev (SYK) models\cite{sachdev1993gapless,
Kitaev:2015tk,
maldacena2016remarks,
fu2016numerical,
kourkoulou2017pure,
huang2017eigenstate}.

To be more concrete, we want to calculate the entanglement entropies for all possible bipartitions of both past and future qudits in the unitary evolution generated by random Hamiltonians. All these data are summarized as what we called the \emph{entanglement features}\cite{you2018machine} of the unitary evolution, which characterizes all the entanglement properties of the corresponding quantum dynamics. An idea that we wish to put forward is to think of the entanglement entropy as a kind of ``free energy'' associated to each configuration of entanglement regions\cite{
hayden2016holographic, qi2017holographic, you2018machine}. The underlying statistical mechanical model that reproduces the free energy functional then provides an efficient description of the entanglement features. Such a statistical mechanical interpretation of quantum many-body entanglement originated in the study of random tensor networks\cite{
hayden2016holographic}, where it was shown that the entanglement entropy of a random tensor network state can indeed be mapped to the free energy of a statistical mechanical model defined on the same graph as the tensor network. The model can be as simple as an Ising model if the 2nd-R\'enyi entropy is considered. 
%The domain wall in the Ising model corresponds to the entanglement cut through the tensor network, which also admits the interpretation as the minimal surface in the holographic bulk geometry. 
It is also shown that the holographic Ising model can be constructed from the entanglement features by machine learning\cite{you2018machine}, which decodes the emergent holographic geometry from quantum entanglement. In this work, we will follow the same idea to reveal the holographic Ising model that describes the entanglement features of the random Hamiltonian dynamics. 
The holographic interpretation provides us a toy model of black hole formation in the holographic bulk under quantum chaotic dynamics on the holographic boundary.

Another practical motivation of this work is to bridge the two existing notions of thermalization in quantum many-body systems: eigenstate thermalization\cite{Deutsch1991, Srednicki1994, Rigol2008,nandkishore2015many} and quantum chaos\cite{
furuya1998quantum,lakshminarayan2001entangling,hosur2016chaos}. The eigenstate thermalization focus on the static (equilibrium) aspects of thermalization, such as the energy level statistics and the reduced density matrix of a single eigenstate.\cite{Chandran:2014ls,
Oganesyan2007,
Atas2013,
BarLev2015,
Santos2010,
you2017sachdev} The quantum chaos focus on the dynamical aspects of thermalization, such as entropy growth, butterfly effect and information scrambling. The  relation between these two notions of thermalization is still under active investigation. A minimal theoretical description for the eigenstate thermalization is the random matrix theory\cite{Bohigas:1984fk, DAlessio:2016mw}, where the quantum many-body Hamiltonian is treated as a random matrix. This relatively crude model already provides nice predictions of many properties of a thermalizing system, including the Wigner-Dyson level statistics and the volume-law entanglement entropy. On the side of quantum chaos, several measures has been proposed to characterize the chaotic dynamics, including the tripartite information,\cite{hosur2016chaos} the out-of-time-order correlation (OTOC)\cite{larkin1969quasiclassical,Kitaev2014hidden,maldacena2016bound,aleiner2016microscopic,caputa2016scrambling,gu2016fractional,bagrets2017power,huang2017out,fan2017out,chen2017out,iyoda2017scrambling,swingle2017slow,slagle2017out,chen2016quantum,li2017measuring,garttner2017measuring,he2017characterizing}, and the entanglement growth after a quantum quench
\cite{vznidarivc2008many,bardarson2012unbounded,
serbyn2013universal}. These measures can be unified and formulated systematically in terms of entanglement features of the unitary evolution itself. Therefore by studying the entanglement features of random Hamiltonian dynamics, we can learn about the typical quantum chaotic behavior of many-body systems that exhibits eigenstate thermalization.

\section{General Discussions}

\subsection{Definition of Entanglement Features}
We consider a quantum many-body system made of $N$ \emph{qudits} (each qudit has a Hilbert space of dimension $d$). The total Hilbert space $\scH$ is the tensor product of all qudit Hilbert spaces, whose dimension is $D=d^N$. A \emph{random Hamiltonian} is a $D\times D$ Hermitian operator $H$ acting on $\scH$ and drawn from a Gaussian unitary ensemble (GUE), described by the following probability density
\eq{P(H)\propto e^{-\frac{D}{2}\Tr H^2}.}
A \emph{random Hamiltonian dynamics} is an unitary time-evolution generated by a fixed (time-independent) GUE Hamiltonian. These unitary operators forms an ensemble that evolves with time
\eq{\label{eq: scE}\scE(t)=\{U(t)=e^{-\ii H t}|H\in\text{GUE}\}.}
The ensemble $\scE(t)$ starts with a simple limit at $t=0$ containing just the identity operator and gradually evolves into a complicated random unitary ensemble (but not exactly Haar-random\cite{cotler2017chaos} in the long time limit) which entangles all qudits together. With the tensor product structure of the Hilbert space, we will be able to address how the entanglement is generated among different subsets of qudits. %Subsystem entanglement entropies of non-local Hamiltonian have also been discussed in the setting of SYK model\cite{fu2016numerical,kourkoulou2017pure,huang2017eigenstate}.

A unitary operator can be graphically represented as
\eq{U(t)={\dia{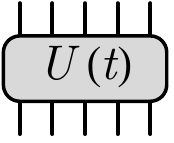}{26}{-10}}_\text{input}^\text{output},}
where each leg represents the action in a qudit Hilbert space and the time flows upwards. It can also be viewed as a \emph{quantum gate}, where the bottom (top) legs are input (output) quantum channels. This tensor-network-like picture encourages us to treat the unitary gate as an (unnormalized) quantum state, such that we can ask about the entanglement entropies of different subsets of the input and output channels.

To describes the entanglement property of the unitary gate $U(t)$ systematically, we introduce the concept of \emph{entanglement features}.\cite{you2018machine}
\begin{defn}[Entanglement feature]
The entanglement features of a unitary gate $U$ refer to the collection of (exponentiated) entanglement entropies over all partitions of the input and output channels to all orders of R\'enyi index. Each specific entanglement feature $W_U^{(n)}[\sigma,\tau]$ is defined as
\eq{\label{eq: def WU}W_U^{(n)}[\sigma,\tau]=\Tr U^{\otimes n}X_\sigma(U^{\otimes n})^\dagger X_\tau,}
where $n$ is the R\'enyi-index and $U^{\otimes n}$ is the $n$-replication of the unitary $U$. 
Given the R\'enyi index $n$, the entanglement feature is specified by two permutation group elements $\sigma,\tau\in S_n^{\times N}$, which can be written in the component form as $\sigma=\sigma_1\times\sigma_2\times\cdots\times\sigma_N$ and similarly for $\tau$. Each element $\sigma_i\in S_n$ represents a permutation among the $n$ replica of the $i$-th qudit. $X_\sigma$ denotes the representation of $\sigma\in S_n^{\times N}$ in the $n$-replicated Hilbert space $\scH^{\otimes n}$. 
\end{defn}
As $U(t)$ evolves in time, its entanglement features also change. In fact, $W_{U(t)}^{(n)}[\sigma,\tau]$ can be considered as the time correlation function $
W_U^{n}[\sigma,\tau] = \Tr X_\sigma (t) X_\tau 
$ between Heisenberg evolved permutations $X_{\sigma}(t)= U^{\otimes n} X_\sigma (U^{\otimes n})^\dagger$ and $X_\tau$ in the replicated Hilbert space $\scH^{\otimes n}$.
The entanglement features are directly related to the entanglement entropies of the unitary gate
\cite{hosur2016chaos,liu2017entropic,liu2018generalized} (by definition)
\eq{\label{eq: S=lnW}S_{U}^{(n)}[\sigma,\tau]=\frac{1}{1-n}\ln \frac{W_{U}^{(n)}[\sigma,\tau]}{D^n}.}
The von Neumann entropy corresponds to the limit that $n\to 1$ by analytic continuation. In this notation, the entanglement region $A$ is specified by the permutations $\sigma$ and $\tau$ according to the assignment of either the cyclic $\cyc$ (like \dia{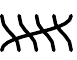}{11}{-2}) or the identity $\id$ (like \dia{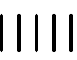}{11}{-2}) permutations,
\eq{\label{eq: A to cyc}\sigma_{i},\tau_{i}=\left\{\begin{array}{cc}\cyc & \text{if }i\in A,\\ \id & \text{if }i\notin A.\end{array}\right.}
Putting together all R\'enyi indices $n$ and all permutations $\sigma$ and $\tau$, the time-dependent entanglement features capture the full information of the entanglement dynamics under the unitary evolution $U(t)$. 

In Appendix~\ref{appendix}, we study the entanglement features of two-qudit gates generated by random Hamiltonian, where we notice that as the qudit dimension $d$ becomes large, the ensemble fluctuation for entanglement features is suppressed quickly. Therefore in the following sections we will focus on the \emph{ensemble averaged} entanglement features
\eq{\label{eq: def W}W^{(n)}[\sigma,\tau]=\langle W_{U}^{(n)}[\sigma,\tau]\rangle_{U\in\scE(t)}.}
The ensemble averaged entanglement feature $W^{(n)}[\sigma,\tau]$ is time-dependent (although not spelt out explicitly), as the unitary ensemble $\scE(t)$ evolves with time according to \eqnref{eq: scE}. 

\subsection{Connections to Other Quantities of Interest}

The entanglement features are useful as they are closely related to many important characteristics of thermalization. For example, the growth of entanglement entropy after a global quench $\ket{\psi(t)}=U(t)\ket{\psi(0)}$ on a product state $\ket{\psi(0)}$ is given by
\eq{\label{eq: Spsi}S_{\ket{\psi(t)}}^{(n)}[\tau]=\frac{1}{1-n}\ln\frac{\Gamma(d)^N}{\Gamma(d+n)^N}\sum_{[\sigma]}W_{U(t)}^{(n)}[\sigma,\tau].}
%\nts{Discuss operator growth, entanglement negativity.}
Also, the operator-averaged OTOC can be expressed in terms of the entanglement features of the unitary\cite{hosur2016chaos,fan2017out}. Consider $A$ and $B$ are two subsets containing $N_A$ and $N_B$ qudits respectively. Let $O_A$ and $O_B$ be Hermitian operators supported on $A$ and $B$, and $O_A(t)=U(t)O_AU^\dagger(t)$ be the time-evolved operator. As we average over all Hermitian operators $O_A$ and $O_B$ within their supports, the OTOC at infinite temperature can be related to the 2nd-R\'enyi entanglement feature $W_{U(t)}^{(2)}$ by
\eqs{\label{eq: def OTOC}
\OTOC(A,B)&\equiv\mathop{\mathrm{avg}}\limits_{O_A,O_B}\frac{1}{D}\Tr O_A(t) O_B O_A(t) O_B\\
&= d^{-N-N_A-N_B} W_{U(t)}^{(2)}[\sigma,\tau],}
given that the permutations $\sigma$ and $\tau$ are determined by the operator supports $A$ and $B$ as
\eq{\label{eq: st=AB}\sigma_{i}=\left\{\begin{array}{cc}\cyc & \text{if }i\in A,\\ \id & \text{if }i\notin A;\end{array}\right.\quad
\tau_{i}=\left\{\begin{array}{cc}\id & \text{if }i\in B,\\ \cyc & \text{if }i\notin B.\end{array}\right.}
Therefore, we can gain much understanding of the random-Hamiltonian-generated quantum chaotic dynamics by studying the entanglement features of the corresponding unitary evolutions. Although the above framework is quite general, calculating all entanglement features is rather difficult. To keep things simple, we will mainly focus on the 2nd-R\'enyi entanglement features (i.e., the $n=2$ case). It turns out that the 2nd-R\'enyi entanglement features are sufficient to capture all the 4-point operator-averaged OTOC as in \eqnref{eq: def OTOC}, which is of our main interest.

\section{Ensemble Averaged Entanglement Features}

\subsection{Spectral Form Factors}
A random Hamiltonian generated unitary evolution $U(t)=e^{-\ii H t}$ can always be diagonalized as
\eq{\label{eq: U=VLV}U(t)=V\Lambda(t) V^\dagger,}
where $V$ is the unitary matrix that also diagonalize the Hamiltonian $H$, and $\Lambda(t)$ is a diagonal matrix whose diagonal elements are phase factors $\Lambda(t)_{mm}=e^{-\ii E_m t}$ specified by the eigen energies $E_m$ of $H$. For random Hamiltonians taken from the GUE, $V$ are simply Haar random unitaries, and the energy levels follow the joint probability distribution
\eq{P_\text{GUE}[E]\propto \prod_{m>m'}(E_{m}-E_{m'})^2e^{-\frac{D}{2}\sum_{m}E_{m}^2}.}
The statistical features of the energy spectrum can be encoded in the \emph{spectral form factors}, generally defined as
\eq{\begin{split}\scR_{[k]}(t)&=\langle e^{-\ii t \sum_{i} k_i E_{m_i}}\rangle_\text{GUE}\\&=\int_{[E]}P_\text{GUE}[E] e^{-\ii t \sum_{i} k_i E_{m_i}},\end{split}}
where $[k]=[k_1,\cdots,k_l]$ is a set of integers $k_i\in\dsZ$ that labels the spectral form factor. $\scR_{[k]}$ is non vanishing only if $[k]$ satisfies the neutralization condition, i.e., $\sum_{i} k_i=0$. Due to the $E\to-E$ symmetry of the GUE distribution, the spectral form factor is even in $[k]$, i.e., $\scR_{[k]}=\scR_{[-k]}$. 
Analytic formulae for some of the spectral form factors can be found in Ref.~\onlinecite{cotler2017chaos}, which are rather complicated and will not be repeated here. Here we would just mention the asymptotic forms to the leading order in $D$, 
\eq{\label{eq: scR large D}\scR_{[k]}(t)=\prod_i\frac{J_1(2k_i t)}{k_i t}+\mathcal{O}(D^{-1}),}
where $J_1$ is the Bessel function of the first kind.

Sometimes, it is convenient to introduce another notation of the spectral form factor, labeled by permutation group elements, which is defined as
\eq{\label{eq: Rg}R_{g}^{(n)}(t)=\frac{1}{\Tr \dsX_g}\langle\Tr (\Lambda(t)^{\otimes n}\otimes \Lambda^*(t)^{\otimes n}) \dsX_g\rangle_\text{GUE},}
where $\Lambda^*(t)=\Lambda(-t)$ is the complex conjugate of the diagonal phase matrix. Both $\Lambda$ and $\Lambda^*$ are $n$-replicated, which leads to totally $2n$ layers. $g\in S_{2n}$ is a permutation among these $2n$ layers and $\dsX_g$ denotes the matrix representation of $g$ in the $\scH^{\otimes 2n}$ Hilbert space. 

\subsection{Ensemble Average}

\begin{figure}
\includegraphics[scale=0.25]{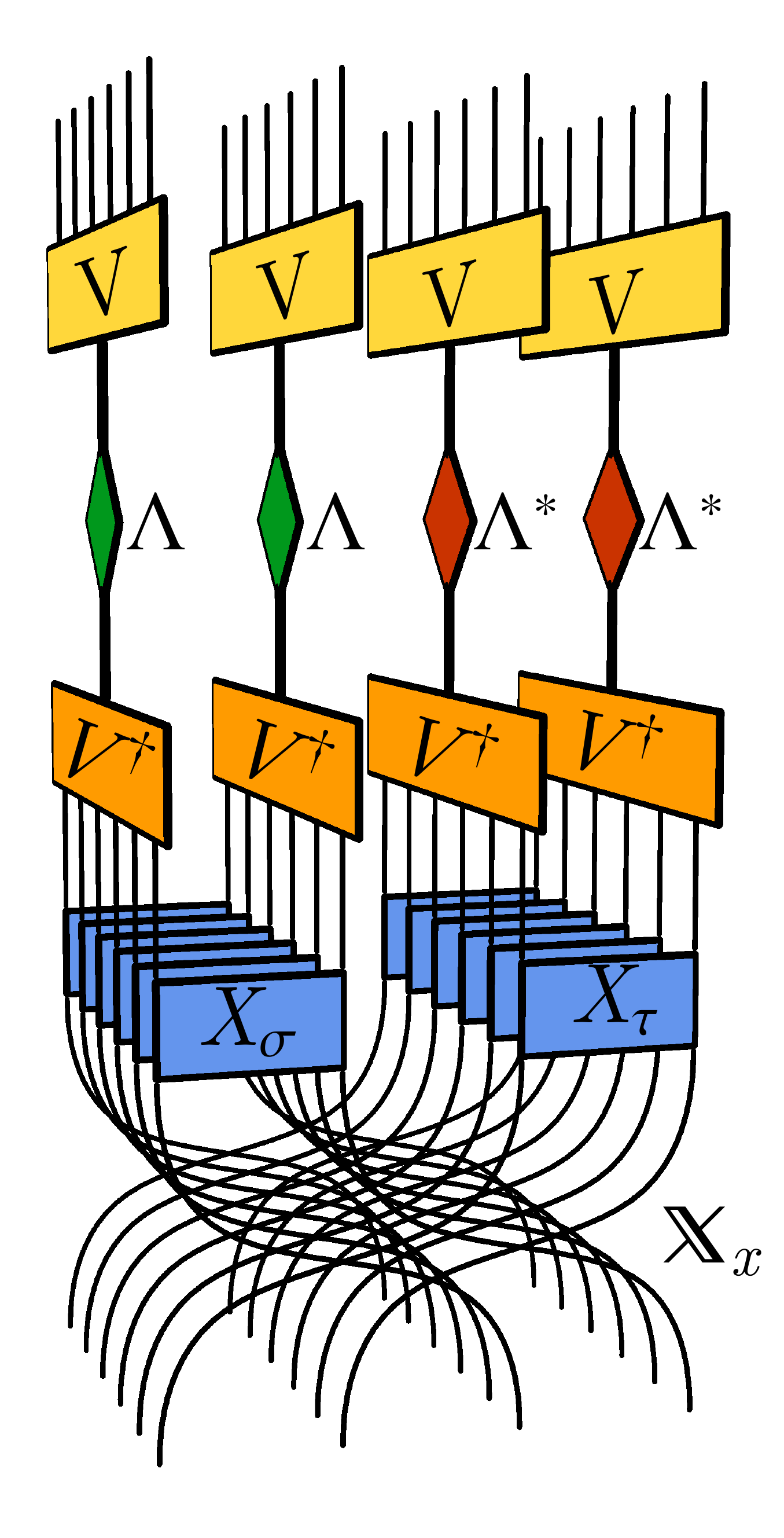}
\caption{The diagrammatic representation of \eqnref{eq: W diag} for the $n=2$ case. The generalization to $n>2$ cases is straightforward.}
\label{fig: diagram}
\end{figure}

Now we are in the position to calculate the ensemble averaged $n$th-R\'enyi  entanglement features defined in \eqnref{eq: def W}.
Plugging in the definition of entanglement feature in \eqnref{eq: def WU} and express the unitary evolution in its diagonal basis following \eqnref{eq: U=VLV}, we can rearrange \eqnref{eq: def W} into
\eq{\label{eq: W diag}\begin{split}
&W^{(n)}[\sigma,\tau]=\langle \Tr (V\Lambda V^\dagger)^{\otimes n}X_\sigma (V\Lambda^* V^\dagger)^{\otimes n}X_\tau\rangle\\
&=\langle\Tr \underbrace{ V^{\otimes 2n}(\Lambda^{\otimes n}\otimes \Lambda^{*\otimes n})V^{\dagger\otimes 2n}(X_\sigma\otimes X_\tau)\dsX_x}_{\text{(see \figref{fig: diagram} for diagrammatic representation)}}\rangle\\
%&=\langle\Tr\dia{W3D}{180}{-90}\rangle.
\end{split}}
We have introduced $\dsX_x$ to represent the large swap operator between the $\Lambda$ layers and the $\Lambda^*$ layers at the bottom of the diagram in \figref{fig: diagram}. The trace operator $\Tr$  acting on the diagram simply connects the top legs to the bottom legs by imposing a ``periodic boundary condition'' in the vertical direction. The ensemble average includes both averaging $V$ and $V^\dagger$ over Haar random unitary ensembles and averaging $\Lambda$ over the energy levels of GUE random matrices.  

Let us first take the Haar ensemble average of $V$ and $V^\dagger$. The result reads\cite{kitaev},
\eq{\label{eq: W=sumWg}\begin{split}
W^{(n)}[\sigma,\tau]=\sum_{g,h\in S_{2n}}&\Wg_{g^{-1}h}\langle\Tr (\Lambda^{\otimes n}\otimes\Lambda^{*\otimes n})\dsX_g\rangle\\
&\Tr \dsX_h (X_\sigma\otimes X_\tau) \dsX_x,
\end{split}}
where $g\in S_{2n}$ stands for permutations among the $2n$ layers, and $\dsX_{g}$ denotes the representation of $g$ in the $\scH^{\otimes 2n}$ Hilbert space. $\Wg_{g}$ is the Weingarten function, which appeared in the integration of Haar random unitaries. $\Wg_{g}$ is a class function in $g$ and is a rational function in the Hilbert space dimension $D$. In the large $D$ limit, the Weingarten function (for $S_{2n}$ group) has the asymptotic form
\eq{\Wg_{g}=D^{-4n+\#(g)}\prod_{i}(-)^{\nu_i(g)-1}C_{\nu_i(g)-1}+\cdots,}
where $\#(g)$ is the number of cycles in $g$ and $\nu_i(g)$ is the length of the $i$th cycle in $g$. $C_m=(2m)!/m!(m+1)!$ is the Catalan number.

We then carry out the ensemble average over energy levels. According to \eqnref{eq: Rg}, the result can be expressed in terms of the spectral form factor as $\langle\Tr (\Lambda^{\otimes n}\otimes\Lambda^{*\otimes n})\dsX_g\rangle=R_g^{(n)}\Tr \dsX_g$. So the problem boils down to evaluating various traces of permutation operators, which is essentially a problem of counting permutation cycles. To evaluate the trace $\Tr\dsX_h (X_\sigma\otimes X_\tau) \dsX_x$, we note that every permutation matrix in this expression is a direct product of the small permutations that acts independently in the quantum channel of each qudit. So the result can be factorized to
\eq{\Tr\dsX_h (X_\sigma\otimes X_\tau) \dsX_x =\prod_{i}d^{-K_h(\sigma_i,\tau_i)},}
where $K_h(\sigma_i,\tau_i)=\#(h(\sigma_i\otimes\tau_i)x)$ is a cycle counting function. 

Putting all pieces together, \eqnref{eq: W=sumWg} becomes
\eq{\label{eq: Wn result}W^{(n)}[\sigma,\tau]=\sum_{g,h\in S_{2n}}\Wg_{g^{-1}h}R_g^{(n)}D^{\#(g)-\overline{K}_h[\sigma,\tau]},}
where $\overline{K}_h[\sigma,\tau]=\frac{1}{N}\sum_i K_h(\sigma_i,\tau_i)$. The time dependence enters from the spectral form factor $R_g^{(n)}(t)$ defined in \eqnref{eq: Rg}. This gives the general formula for the $n$th-R\'enyi entanglement feature averaged over the ensemble $\scE(t)$. In the following, we will restrict to the $n=2$ case and discuss several applications.

\section{2nd R\'enyi Entanglement Features}

\subsection{Ising Formulation}

For $n=2$, \eqnref{eq: Wn result} reduces to
\eq{\label{eq: W2 result}W^{(2)}[\sigma,\tau]=\sum_{g,h\in S_4}\Wg_{g^{-1}h}R_g^{(2)}D^{\#(g)-\overline{K}_h[\sigma,\tau]}.}
where $\sigma=\sigma_1\times\sigma_2\times\cdots\times\sigma_N$ and similar for $\tau$. Here $\sigma_i, \tau_i\in S_2$ are identity or swap operators. But it will be more intuitive to treat them as \emph{Ising variables} living on the input and output channels of the unitary gate respectively, and think of $K_h(\sigma_i,\tau_i)$ as an energy functional that describes the Ising couplings between them. In this regard, we will assign $\pm1$ values to the $S_2(=\mathbb{Z}_2)$ group element as
\eq{\sigma_i,\tau_i=\bigg\{\begin{array}{cc}+1 & \text{for }\dia{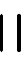}{11}{-2},\\-1 & \text{for }\dia{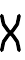}{11}{-2}.\end{array}}
Then the energy functional $K_h(\sigma_i,\tau_i)$ can be enumerated as in \tabref{tab: Kh} for all $h\in S_4$. To evaluate \eqnref{eq: W2 result} we also need to know the spectral form factor $R_g^{(2)}$ for all $g\in S_4$. We can first express $R_g^{(n)}(t)$ in terms of $\scR_{[k]}(t)$. Their correspondences are listed in \tabref{tab: SFF}. These spectral form factors $\scR_{[k]}(t)$ are calculated in Ref.~\onlinecite{cotler2017chaos}, whose notation is differed from ours by a factor of $D$ to some power, see the  last column of \tabref{tab: SFF}.

\begin{table}[htbp]
\caption{The Ising coupling energy $K_h(\sigma_i,\tau_i)$ for different permuations $h\in S_4$.}
\begin{center}
\begin{tabular}{cc}
$h\in S_4$ & $K_h(\sigma_i,\tau_i)$\\
\hline
\dia{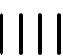}{11}{-2}, \dia{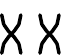}{11}{-2} & $-\frac{1}{2} \sigma _i \tau _i-\frac{3}{2}$\\
\dia{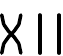}{11}{-2}, \dia{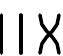}{11}{-2} & $+\frac{1}{2}\sigma _i \tau _i-\frac{3}{2}$\\
\dia{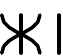}{11}{-2}, \dia{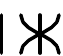}{11}{-2}, \dia{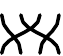}{11}{-2}, \dia{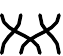}{11}{-2} & $-\frac{1}{2}\sigma _i-\frac{1}{2}\tau _i-2$\\
\dia{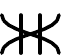}{11}{-2}, \dia{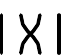}{11}{-2}, \dia{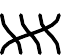}{11}{-2}, \dia{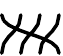}{11}{-2} & $+\frac{1}{2}\sigma _i+\frac{1}{2}\tau _i-2$\\
\dia{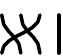}{11}{-2}, \dia{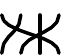}{11}{-2}, \dia{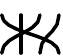}{11}{-2}, \dia{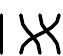}{11}{-2} & $-\frac{1}{2}\sigma _i+\frac{1}{2}\tau _i-2$\\
\dia{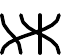}{11}{-2}, \dia{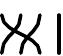}{11}{-2}, \dia{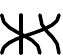}{11}{-2}, \dia{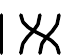}{11}{-2} & $+\frac{1}{2}\sigma _i-\frac{1}{2}\tau _i-2$\\
\dia{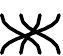}{11}{-2} & $+\frac{1}{2}\sigma _i-\frac{1}{2}\tau _i-3$\\
\dia{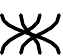}{11}{-2} & $-\frac{1}{2}\sigma _i+\frac{1}{2}\tau _i-3$\\
\dia{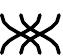}{11}{-2} & $-\frac{1}{2}\sigma _i-\frac{1}{2}\tau _i-3$\\
\dia{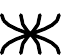}{11}{-2} & $+\frac{1}{2}\sigma _i+\frac{1}{2}\tau _i-3$\end{tabular}
\end{center}
\label{tab: Kh}
\end{table}

\begin{table}[htbp]
\caption{Spectral form factors $R_g^{(2)}$ for different permutations $g\in S_4$ in terms of $\scR_{[k]}$. The last column shows the corresponding notation in Ref.~\onlinecite{cotler2017chaos}.}
\begin{center}
\begin{tabular}{ccc}
$g\in S_4$ & $R_g^{(2)}$ & Ref.~\onlinecite{cotler2017chaos}\\
\hline
\dia{00}{11}{-2} & $\scR_{[11\bar{1}\bar{1}]}$ & $\frac{1}{D^4}\scR_{4}$ \\
\dia{12,34}{11}{-2} & $\scR_{[2\bar{2}]}$ & $\frac{1}{D^2}\scR_{4,2}$\\
\dia{12}{11}{-2}, \dia{34}{11}{-2} & $\scR_{[2\bar{1}\bar{1}]}$ & $\frac{1}{D^3}\scR_{4,1}$\\
\dia{13,24}{11}{-2}, \dia{14,23}{11}{-2} & $\scR_{[00]}$ & \\
\dia{13}{11}{-2}, \dia{14}{11}{-2}, \dia{23}{11}{-2}, \dia{24}{11}{-2} & $\scR_{[1\bar{1}0]}$ & $\frac{1}{D^2}\scR_{2}$\\
\dia{1234}{11}{-2}, \dia{1243}{11}{-2}, \dia{1324}{11}{-2}, \dia{1342}{11}{-2}, \dia{1423}{11}{-2}, \dia{1432}{11}{-2} & $\scR_{[0]}$ & \\
\dia{123}{11}{-2}, \dia{124}{11}{-2}, \dia{132}{11}{-2}, \dia{134}{11}{-2}, \dia{142}{11}{-2}, \dia{143}{11}{-2}, \dia{234}{11}{-2}, \dia{243}{11}{-2} & $\scR_{[1\bar{1}]}$ & $\frac{1}{D^2}\scR_{2}$\\
\end{tabular}
\end{center}
\label{tab: SFF}
\end{table}

No approximation has been made up to this point. If we substitute the exact expressions of both the Weingarten function $\Wg_{g^{-1}h}$ and the spectral form factor $R_g^{(2)}$ to \eqnref{eq: W2 result} and carry out the double summation over the $S_4$ group, we can arrive at the exact result of the ensemble averaged entanglement features $W^{(2)}$. However, the expression is rather complicated to present here (see Appendix \ref{sec:exact} for the full expression), so we will just show the result to the leading order in $D=d^N$,
\eq{\label{eq: W2 large D}\begin{split}
W^{(2)}&[\sigma,\tau]= \scR_{[11\bar{1}\bar{1}]}D^{\frac{3+\overline{\sigma\tau}}{2}}\\
&-2(\scR_{[11\bar{1}\bar{1}]}-\scR_{[2\bar{1}\bar{1}]})D^{\frac{1-\overline{\sigma\tau}}{2}}\\
&+(\scR_{[00]}-\scR_{[11\bar{1}\bar{1}]})(D^{\frac{2+\overline{\sigma}+\overline{\tau}}{2}}+D^{\frac{2-\overline{\sigma}-\overline{\tau}}{2}})\\
&-(2\scR_{[00]}-\scR_{[0]}+2\scR_{[2\bar{1}\bar{1}]}-3\scR_{[11\bar{1}\bar{1}]})\\
&\hspace{12pt}\times(D^{\frac{\overline{\sigma}-\overline{\tau}}{2}}+D^{\frac{-\overline{\sigma}+\overline{\tau}}{2}})+\cdots,
\end{split}}
where the $\overline{\sigma}$, $\overline{\tau}$ and $\overline{\sigma\tau}$ are respectively the mean magnetizations on both input and output sides and the Ising correlation across the unitary gate,
\eq{\label{eq: means}\overline{\sigma}=\frac{1}{N}\sum_i\sigma_i,\quad\overline{\tau}=\frac{1}{N}\sum_i\tau_i, \quad\overline{\sigma\tau}=\frac{1}{N}\sum_i\sigma_i\tau_i.}
Therefore to the leading order in $D$, the 2nd-R\`enyi entanglement features $W^{(2)}$ of random Hamiltonian dynamics can be given by \eqnref{eq: W2 large D} as  Boltzmann weights (partition weights) of the Ising variables $\sigma$ and $\tau$. The time dependence of the the entanglement features are captured by the spectral form factors $\scR_{[k]}$, whose large-$D$ asymptotic behavior was given by \eqnref{eq: scR large D}. Based on this result, we can further explore the entanglement growth and the OTOC under random Hamiltonian dynamics.

\subsection{Holographic Interpretations}

Given $W^{(2)}[\sigma,\tau]$ in the form of a Boltzmann weight, we would like to understand, what kind of Ising model does $W^{(2)}[\sigma,\tau]$ describe? The most naive approach is to follow the standard idea of statistical mechanics and assume that there is a single Ising Hamiltonian $H[\sigma,\tau]$ that models the Boltzmann weight via $W^{(2)}[\sigma,\tau]\propto e^{-H[\sigma,\tau]}$. Such a Hamiltonian would necessarily involve multi-spin interactions in the general form of $H[\sigma,\tau]=\sum J_{i_1\cdots i_m}^{j_1\cdots j_n}\sigma_{i_1}\cdots\sigma_{i_m}\tau_{j_1}\cdots\tau_{j_n}$, which requires exponentially (in $N$) many couplings to parameterize. This naive approach does not provide us a more intuitive understanding of the entanglement features.

How to efficiently represent the ``big data'' of entanglement features? An idea developed in the machine learning community is to encode the exponential amount of data in the polynomial amount of neural network parameters, if the data has strong internal correlations. In this approach, hidden neurons are introduced in the neural network to mediate the many-body correlations among the visible neurons. We will take the similar philosophy to model the entanglement features as a superposition of several Ising models with hidden variables, such that each Ising model only contains  few-body interactions that can be efficiently parameterized by polynomial amount of couplings.

As can be seen from \eqnref{eq: W2 large D}, there are four terms in $W^{(2)}[\sigma,\tau]$, each term can be interpreted as an Ising model with at most two body interactions. Putting these terms together is like statistical superposition of different Ising models defined on different background geometries (graph connectivities) with weights that are not necessarily positive. The superposition of Ising models can be considered as a kind of gravitational fluctuation, as the lattice structure (graph connectivity) of the Ising model is changing from model to model. On each fixed background, the Ising variables have no (connected) correlation beyond two-body. But once the gravitational fluctuations are introduced, complicated many-body correlations will be generated among all Ising variables.

\begin{figure}[t]
\begin{center}
\includegraphics[width=0.95\linewidth]{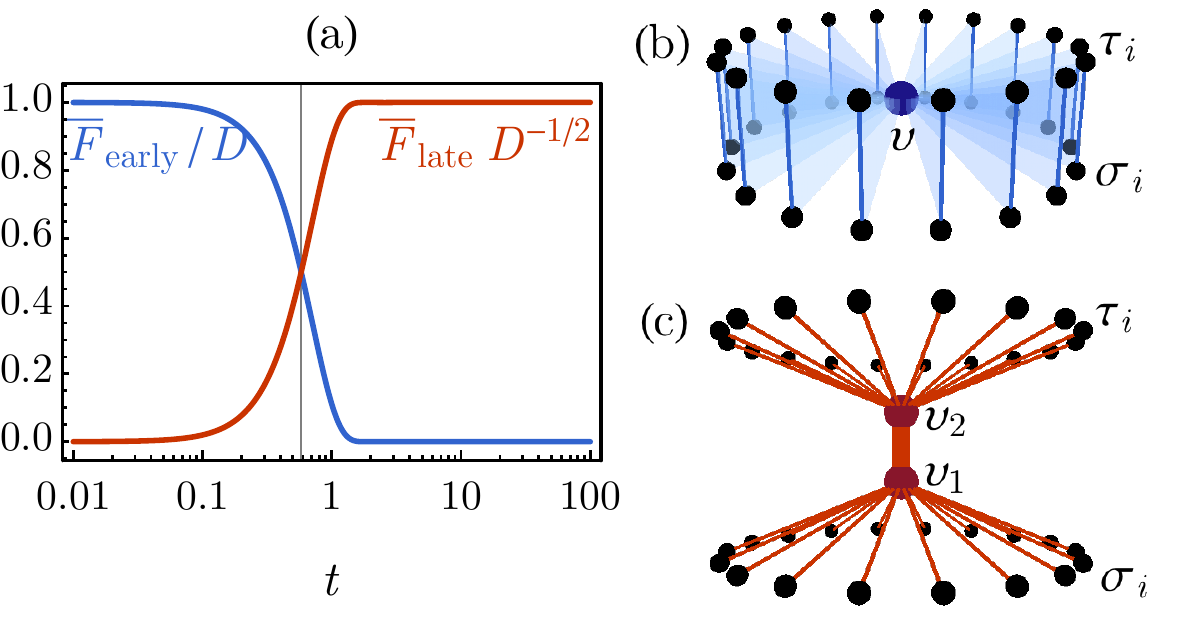}
\caption{(a) The weight functions $\bar{F}_\text{early}$ and $\bar{F}_\text{late}$ v.s. time $t$, showing the crossover from the early-time model $W_\text{early}$ to the late-time model $W_\text{late}$. Holographic Ising models in (b) the early time $W_\text{early}$ and (c) the late time $W_\text{late}$. Each dot is an Ising variable and each bond corresponds to a ferromagnetic Ising coupling. The light triangles in (b) denotes the three-body interaction between boundary and bulk.}
\label{fig: holography}
\end{center}
\end{figure}

If we introduce some auxiliary degrees of freedom in the  holographic bulk, we can separate the entanglement features in \eqnref{eq: W2 large D} into two terms,
\eq{\label{eq: W=W+W}W^{(2)}[\sigma,\tau]=W_\text{early}[\sigma,\tau]+W_\text{late}[\sigma,\tau],}
where $W_\text{early}$ governs the early-time behavior and $W_\text{late}$ governs the late-time behavior (as to be justified soon)
\eq{\label{eq: W=DF}\begin{split}
&W_\text{early}[\sigma,\tau]=\sum_{\upsilon=\pm1}D^{\frac{1}{2}(\upsilon\overline{\sigma\tau}+\upsilon)}F_\text{early}(\upsilon),\\
&W_\text{late}[\sigma,\tau]=\sum_{\upsilon_{1,2}=\pm1}D^{\frac{1}{2}(\upsilon_1\overline{\sigma}+\upsilon_2\overline{\tau}+\upsilon_1\upsilon_2)}F_\text{late}(\upsilon_1\upsilon_2).
\end{split}}
Auxiliary Ising variables $\upsilon$ (or $\upsilon_{1,2}$) are introduced as the bulk degrees of freedom, whose fluctuations are governed by the partition weight $F_\text{early}(\upsilon)$ (or $F_\text{late}(\upsilon_1\upsilon_2)$), which can be directly read off from \eqnref{eq: W2 large D},
\eq{\begin{split}
F_\text{early}(\upsilon)&=\left\{\begin{array}{ll} \scR_{[11\bar{1}\bar{1}]}D & \upsilon=+1,\\
-2(\scR_{[11\bar{1}\bar{1}]}-\scR_{[2\bar{1}\bar{1}]})D & \upsilon = -1;\end{array}\right.\\
F_\text{late}(\upsilon)&=\left\{\begin{array}{ll} (\scR_{[00]}-\scR_{[11\bar{1}\bar{1}]})D^{\frac{1}{2}} & \upsilon=+1,\\
-(2\scR_{[00]}-\scR_{[0]}+2\scR_{[2\bar{1}\bar{1}]} &\\
\hspace{12pt}-3\scR_{[11\bar{1}\bar{1}]})D^{\frac{1}{2}} & \upsilon = -1.\end{array}\right.
\end{split}}
If we trace out the boundary freedoms $\sigma$ and $\tau$, we can obtain the effective theory for the bulk freedom $\upsilon$, from which we can evaluate the expectation value of the weight functions $\bar{F}_\text{early,late}=\langle F_\text{early,late}(\upsilon)\rangle_{\upsilon}$. They characterize the relative importance between the two models $W_\text{early}$ and $W_\text{late}$. We plot $\bar{F}_\text{early,late}(t)$ as a function of time $t$ in \figref{fig: holography}(a). There is a crossover between $\bar{F}_\text{early}$ and $\bar{F}_\text{late}$ around an order-one time scale $t_\text{c}\approx 0.58$ (in unit of the inverse of the energy scale of the GUE Hamiltonian). So the early (late) time entanglement features are indeed dominated by $W_\text{early}$ ($W_\text{late}$).

In the early time, the entanglement features are dominated by $W_\text{early}=\Tr_\upsilon e^{-H_\text{early}}F_\text{early}$, which can be described by an Ising Hamiltonian with a bulk variable $\upsilon$,
\eq{H_\text{early}[\sigma,\tau;\upsilon]=-\frac{\ln d}{2} \sum_{i}\upsilon\sigma_i\tau_i-\frac{\ln D}{2}\upsilon.}
The last term is a strong Zeeman field that pins the bulk variable to $\upsilon=+1$. Then the first term simply describes a direct coupling between input and output along each quantum channels separately, as illustrated in \figref{fig: holography}(b). This is indeed the entanglement feature expected for the unitary gate close to identity. In most cases, the feedback effect from the first term will not be able to overturn the strong Zeeman pinning of the second term, unless the input and output variables are anti-polarized, i.e., $\overline{\sigma\tau}\simeq -1$, which corresponds to choosing the entanglement region to be either the input or the output channels only. In this case, the bulk fluctuates strongly and the entanglement features are dominated by $1/D$ effects. Apart from this strong fluctuation limit, the bulk will be well behaved and the corresponding holographic geometry is a fragmented space (i.e. each quantum channels are far separated from each other in the holographic space, because there is almost no entanglement among them).

In the late time, the entanglement features are dominated by $W_\text{late}=\Tr_{[\upsilon]} e^{-H_\text{late}}F_\text{late}$, which can be described by an Ising Hamiltonian with two bulk variables $\upsilon=[\upsilon_1,\upsilon_2]$,
\eq{H_\text{late}[\sigma,\tau;\upsilon]=-\frac{\ln d}{2}\sum_{i}(\upsilon_1\sigma_i+\upsilon_2\tau_i)-\frac{\ln D}{2}\upsilon_1\upsilon_2.}
The late-time model only contains two-body interactions as illustrated in \figref{fig: holography}(c). All the input (output) variables couples to $\upsilon_1$ ($\upsilon_2$) with coupling strength $\ln d/2$. The bulk variables $\upsilon_1$ and $\upsilon_2$ themselves couples strongly with the strength $\ln D/2$ (which is $N$ times stronger than $\ln d/2$). As shown in Ref.~\onlinecite{hayden2016holographic}, the holographic Ising model implies to a random tensor network description (of the unitary gate) with the same network geometry. In the tensor network description, all quantum information from the input side enters the tensor $\upsilon_1$ gets scrambled. The scrambled information are then emitted from the tensor $\upsilon_2$ to the output side. This implies that $\upsilon_1$ and $\upsilon_2$ can be considered as a pair of temporally entangled black hole and white hole in the holographic bulk, matching the holographic interpretation of quantum chaotic unitary evolution in the late-time regime.

\subsection{Early-Time and Late-Time Limits}

In this section, we will go beyond the leading $D$ result in \eqnref{eq: W2 large D} and list some exact results of ensemble averaged entanglement features in the early-time and late-time limits. In the early-time limit, the unitary ensemble $\scE(t=0)$ contains only the identity gate, whose 2nd-R\'entyi entanglement features are given by
\eq{W_0^{(2)}[\sigma,\tau]=D^{\frac{3+\overline{\sigma\tau}}{2}}.}
which can be derived from \eqnref{eq: W2 result} by using the fact that $R_g^{(2)}(t=0)=1$ for any $g\in S_4$.

In the late-time limit, the unitary ensemble $\scE(t\to\infty)$ approaches to a random unitary ensemble but not exactly  Haar random. The deviation from the Haar random unitary ensemble has to do with the non-vanishing late-time limit of the following spectral form factors,\cite{cotler2017chaos}
\eqs{&\scR_{[11\bar{1}\bar{1}]}=\frac{2D-1}{D^3},\scR_{[2\bar{1}\bar{1}]}=\frac{1}{D^2},\\
&\scR_{[2\bar{2}]}=\scR_{[1\bar{1}0]}=\scR_{[1\bar{1}]}=\frac{1}{D}.}
Using these late-time limit of the spectral form factors and evaluate \eqnref{eq: W2 result}, we can obtain the late-time limit ($t\to\infty$) of the 2nd-R\'eny entanglement feature (to all order of $D$):
%\eqs{W_\infty^{(2)}[\sigma,\tau]&=\sum_{\upsilon=\pm1}D^{\frac{1}{2}(\upsilon\overline{\sigma\tau}+\upsilon)}F_1(\upsilon)\\&+\sum_{\upsilon_1,\upsilon_2=\pm1}D^{\frac{1}{2}(\upsilon_1\overline{\sigma}+\upsilon_2\overline{\tau}+\upsilon_1\upsilon_2)}F_2(\upsilon_1\upsilon_2).}
\eqs{\label{eq: W2 late}W_\infty^{(2)}[\sigma,\tau]&=\frac{1}{(D+1)(D+3)}\big(2D^{\frac{1}{2}}((D+2)D^{\frac{\overline{\sigma\tau}}{2}}-D^{-\frac{\overline{\sigma\tau}}{2}})\\
&+D(D^2+4D+2)(D^{\frac{\overline{\sigma}+\overline{\tau}}{2}}+D^{-\frac{\overline{\sigma}+\overline{\tau}}{2}})\\
&-D(D+4)(D^{\frac{\overline{\sigma}-\overline{\tau}}{2}}+D^{-\frac{\overline{\sigma}-\overline{\tau}}{2}})\big),}
where $\overline{\sigma}$, $\overline{\tau}$, $\overline{\sigma\tau}$ were defined in \eqnref{eq: means}. In comparison, the 2nd-R\'enyi entanglement features of Haar random unitaries are given by
\eqs{\label{eq: W2 Haar}W_\text{Haar}^{(2)}[\sigma,\tau]&=\frac{D^2}{D^2-1}\big(D(D^{\frac{\overline{\sigma}+\overline{\tau}}{2}}+D^{-\frac{\overline{\sigma}+\overline{\tau}}{2}})\\
&-(D^{\frac{\overline{\sigma}-\overline{\tau}}{2}}+D^{-\frac{\overline{\sigma}-\overline{\tau}}{2}})\big).}
$W_\infty^{(2)}$ and $W_\text{Haar}^{(2)}$ have the same large-$D$ limit. Their difference is revealed only at the sub-leading order of $D$.

\section{Applications and Numerics}

\subsection{Input-Output Mutual Information}

\begin{figure}[t]
\begin{center}
\includegraphics[width=0.88\linewidth]{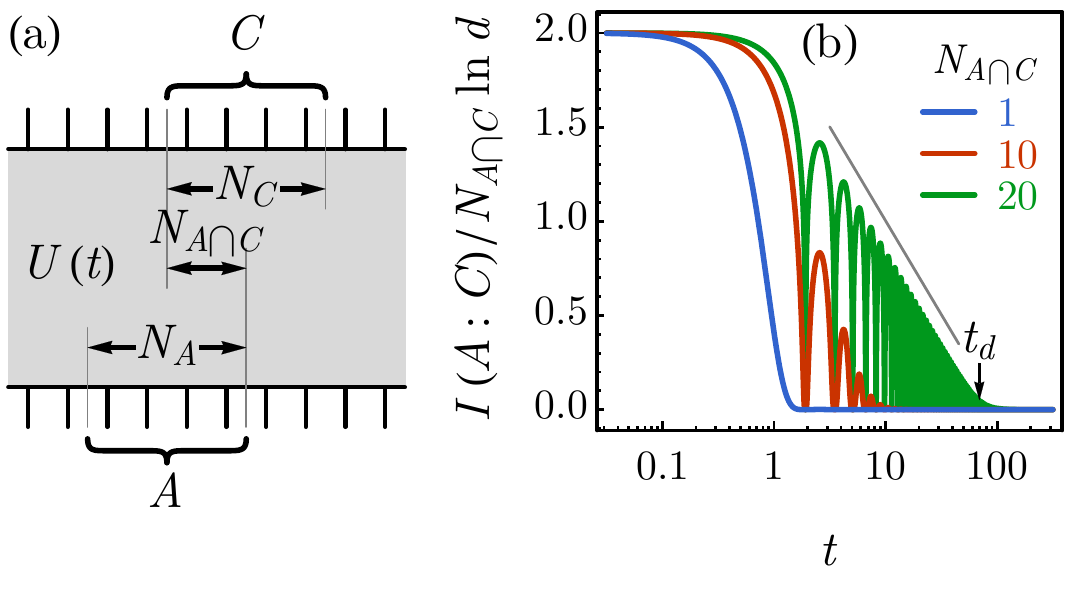}
\caption{(a) Specifications of subsets $A$ and $C$ on the input and output sides of the unitary gate $U(t)$. (b) Mutual information $I(A:C)$ in unit of $N_{A\cap C}$ dits (take $d=2$ for instance here). Different colors correspond to different $N_{A\cap C}$. For large $N_{A\cap C}$, the envelop decays with $\ln t$ linearly.}
\label{fig: IAC}
\end{center}
\end{figure}

As an application of the entanglement features, let us first consider the mutual information between input and output channels for $N$ qudits. Suppose $N$ is large (and correspondingly $D=d^N$ is large), we can use the leading $D$ result in \eqnref{eq: W2 large D} to analyze the entanglement properties. Consider a subset $A$ of input channels and a subset $C$ of output channels of the unitary gate $U\in\scE(t)$, we are interested in the ensemble-averaged 2nd-R\'enyi mutual information $I(A:C)=S^{(2)}(A)+S^{(2)}(C)-S^{(2)}(AC)$ between subregions $A$ and $C$. Let $N_A$ (or $N_C$) be the number of qudits in $A$ (or $C$), and $N_{A\cap C}$ (or $N_{A\cup C}$) be the number of qudits in the intersection (or union) of $A$ and $C$, as illustrated in \figref{fig: IAC}(a). With this setup, $S^{(2)}(A)=N_A\ln d$ and $S^{(2)}(C)=N_C\ln d$ are trivially determined, and $S^{(2)}(AC)=-\ln W^{(2)}[\sigma,\tau]/D^2$ can be expressed in terms of the entanglement feature with the Ising variables following $\overline{\sigma}=1-2N_A/N$, $\overline{\tau}=1-2N_C/N$, $\overline{\sigma\tau}=1-2(N_{A\cup C}-N_{A\cap C})/N$. 

Using the result in \eqnref{eq: W2 large D}, for $d^{N_A+N_C}\ll d^N$, the mutual information between the input subregion $A$ and the output subregion $C$ follows
\eq{\label{eq: IAC} I(A:C)=\ln\big(\scR_{[11\bar{1}\bar{1}]}(d^{2N_{A\cap C}}-1)+1\big),}
where the spectral form factor simply reads $\scR_{[11\bar{1}\bar{1}]}=(J_1(2t)/t)^4$ as the large-$D$ limit is taken. As time evolves, the mutual information $I(A:C)$ deveats from $2N_{A\cap C}\ln d$ to nearly zero $\scO(D^{-2})$, as shown in \figref{fig: IAC}(b) for different $N_{A\cap C}$. This describes how the input and output qudits lose mutual information under random Hamiltonian dynamics as a result of the information scrambling.  The time that $I(A:C)$ first approaches $0$ is of the same order as the scrambling time of this system, which is an $\scO(1)$ time scale. For large $N_{A\cap C}$, it takes exponentially long time for $I(A:C)$ to approach zero, as demonstrated in \figref{fig: IAC}(b). This time scale can be identified as the ``dip time''  $t_d$ introduced in Ref.~\onlinecite{cotler2017black}, which is set by the equation $\scR_{[11\bar{1}\bar{1}]}(t_d)d^{2N_{A\cap C}}\sim 1$. For large $N_{A\cap C}$,
\eq{\label{eq: tstar}t_d=(d^{N_{A\cap C}}/\pi)^{1/3}.}
Within the intermediate time range {$1\lesssim t \lesssim t_d$}, the mutual information $I(A:C)$ decays linearly with $\ln t$,
\eq{I(A:C)\simeq 2N_{A\cap C}\ln d-6\ln t-2\ln\pi,}
as seen in \figref{fig: IAC}(b).

\begin{figure}[t]
\begin{center}
\includegraphics[width=0.9\linewidth]{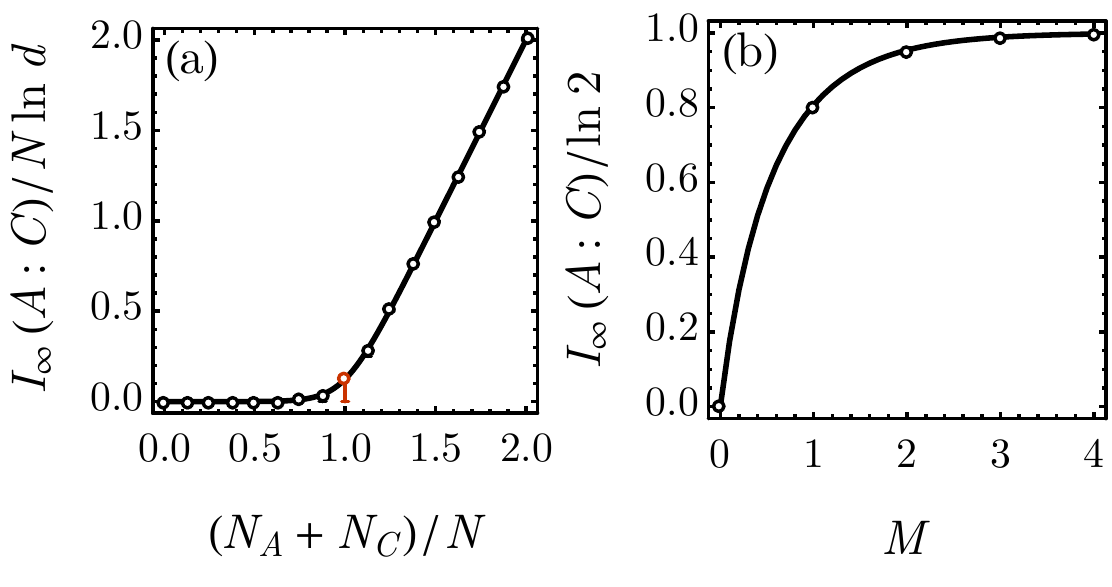}
\caption{The late-time mutual information $I_\infty(A:C)$ in a $d=2$, $N=8$ system. (a) For generic $N_A$ and $N_C$, the numerically calculated $I_\infty(A:C)$ data points (small circles) fall along the theoretical curve of \eqnref{eq: Iinf AC}. The largest deviation is at $N_A+N_C=N$ (red circle with large error bar). In that case, another setup $N_A=M$, $N_C=N-M$ is considered in (b) to expose the deviation from \eqnref{eq: Iinf AC} at small $M$, where the numerical data points (small circles) coincide with the curve of \eqnref{eq: Iinf M}.}
\label{fig: Iinf}
\end{center}
\end{figure}

The late-time ($t>t_d$) saturation value of $I(A:C)$ can be calculated from \eqnref{eq: W2 late}. The result is
\eqs{\label{eq: Iinf}I_\infty(A:C)&=\ln\bigg(\frac{1}{(D+1)(D+3)}\Big(2(1+2D^{-1})d^{2N_{A\cap C}}\\
+&(D^2+4D+2)\big(1+D^{-2}d^{2(N_A+N_C)}\big)\\
-&(1+4D^{-1})\big(d^{2N_A}+d^{2N_C}\big)-2D^{-2}d^{2N_{A\cup C}}\Big)\bigg).}
This is exact to all orders of $D$, but looks rather complicated. In the following, we consider two limits where $I_\infty(A:C)$ admits a simpler (approximate) expression. One limit is the large-$D$ limit, when \eqnref{eq: Iinf} is dominated by its second line and can be approximated by
\eq{\label{eq: Iinf AC} I_\infty(A:C)\simeq \ln\big(1+d^{2(N_A+N_C-N)}\big).}
The approximation works well unless $|N_A-N_C|\to N$. In that limit, we can consider a special (but useful) case of $N_A=M$, $N_C=N-M$ with $d^M\ll d^N$, then
\eq{\label{eq: Iinf M} I_\infty(A:C)\simeq \ln\big(2-d^{-2M}\big).}
They can be benchmarked with numerics, as shown in \figref{fig: Iinf}, where the numerical calculation of $I_\infty(A:C)$ is performed for eight-qubit ($N=8$, $d=2$) random Hamiltonians and the result validates \eqnref{eq: Iinf AC} and \eqnref{eq: Iinf M}. By comparing early-time \eqnref{eq: IAC} and late-time \eqnref{eq: Iinf AC} formulae, we can see that $I(A:C)$ is mainly a function of $N_{A\cap C}$ in the early time, which crosses over to a function of $N_A+N_C$ in the late time. This behavior is associated to the crossover of the entanglement features $W^{(2)}$ as analyzed previously.

\begin{figure}[t]
\begin{center}
\includegraphics[width=0.95\linewidth]{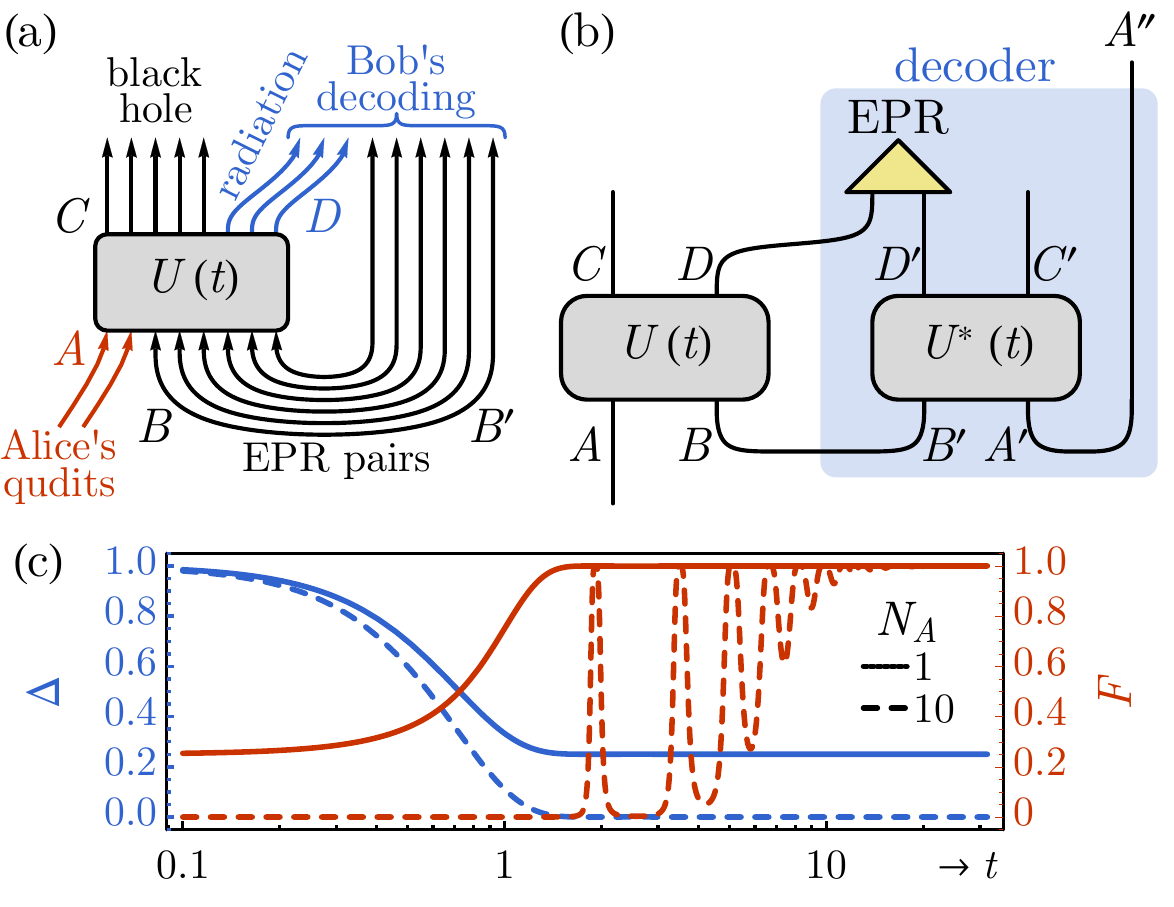}
\caption{(a) The Hayden-Preskill problem. (b) The Yoshida-Kitaev protocol for a probabilistic decoder. (c) The success rate $\Delta$ (in blue) of the Yoshida-Kitaev probabilistic decoder and its fidelity $F$ (in red), calculated for qubit ($d=2$) systems under the assumption of $d^{N_A}\ll d^{N_D}\ll d^N$. The solid (or dashed) curves correspond to the case of $N_A=1$ (or $N_A=10$).}
\label{fig: HP}
\end{center}
\end{figure}

As a fun application, we can implement our result to analyze the Hayden-Preskill problem,\cite{hayden2007black} as illustrated in \figref{fig: HP}(a). The problem can be formulated as follows. Alice has some qudits encoding some confidential quantum information. She throws her qudits $A$ into a black hole $B$ hoping to hide the information forever. But Bob is spying on Alice. He has a system $B'$ which was maximally entangled with the black hole $B$  before Alice threw her qudits in. Then Bob captures some Hawking radiation $D$ at time $t$ after Alice's qudits have been thrown in. Can Bob recover the quantum information about Alice's qudits from the captured radiation $D$ and the purifying system $B'$?  Hayden and Preskill showed that the decoding task is information-theoretically possible. 

Recently, Yoshida and Kitaev proposed a probabilistic and a deterministic decoder in Ref.~\onlinecite{yoshida2017efficient}. In probabilistic protocol, as illustrated in \figref{fig: HP}(b), Bob first prepare a maximally entangled state between $A'$ and $A''$, each contains the same amount $N_A$ of qudits as Alice's qudits $A$. Then he applies the unitary gate $U^*(t)$ to $A' B'$, where $U^*(t)$ is the complex conjugate of the unitary evolution $U(t)$ of the black hole and Alice's qudits. Essentially, Bob needs a quantum computer to simulate the quantum evolution $U^*(t)$. Then he projects the captured Hawking radiation $D$ and its counterpart $D'$ onto the standard EPR state. The projection will either succeed with probability
\eq{\label{eq: Delta}\Delta=\frac{1}{d^{2 N_A}}e^{I(A:C)},}
or signal a failure with probability $1-\Delta$. If succeeded,
the projection has the effect of ``teleporting'' Alice's qudits $A$ to Bob's qudits $A''$ with a fidelity of
\eq{\label{eq: F}F=e^{-I(A:C)}.}
The decoding becomes possible if the black hole has effectively forgotten Alice's information, i.e. $I(A:C)\to 0$ (and the fidelity $F\to 1$ correspondingly). As $I(A:C)$ approaches to zero with time $t$, Alice's information becomes available to Bob in the encoded form, which can be in principle recovered from the Hawking radiation $D$ and purifying system $B'$.

We assume that the time-evolution of the black hole can be modeled by a random Hamiltonian dynamics, then we can apply our result to study the time-dependence of the protocol success rate $\Delta$ and the decoder fidelity $F$. First of all, in the late time limit $t\to\infty$, we can apply \eqnref{eq: Iinf AC} to obtain the saturation values of $\Delta$ and $F$,
\eqs{\Delta_\infty&=\frac{1}{d^{2N_A}}+\frac{1}{d^{2N_D}},\\
F_\infty&=\frac{1}{1+d^{2(N_A-N_D)}}.}
As long as $d^{N_D}\gg d^{N_A}$, the fidelity will be close to one. Therefore to decode Alice's quantum information of $N_A$ qudits, Bob only need to collect a few more qudits ($N_D>N_A$) from the Hawking radiation.

In the case of $d^{N_A}\ll d^{N_D}\ll d^N$, the mutual information $I(A:C)$ can be evaluated using \eqnref{eq: IAC} (where typically $N_{A\cap C}\simeq N_A N_C/N\to N_A$), then from \eqnref{eq: Delta} and \eqnref{eq: F} we can obtain $\Delta$ and $F$ as functions of the radiation capture time $t$, as plotted in \figref{fig: HP}(c). For larger $N_A$, it takes exponentially long time $t_d=(d^{N_A}/\pi)^{1/3}$ for the information of Alice's qudits to be fully scrambled. But Bob does not need to wait for such a long time, because there are a sequence of time windows in the intermediate time regime {$1\lesssim t\lesssim t_d$} in which the fidelity can approach to one shortly, as demonstrated by the $N_A=10$ case in \figref{fig: HP}(c). So it is possible to recover the information of Alice's qudits from these intermediate-time radiations, if Bob can seize the moment.

\subsection{Out-of-Time-Order Correlation}

Now we turn to the operator-averaged OTOC under random Hamiltonian dynamics. As previously defined in \eqnref{eq: def OTOC}, we use $\OTOC(A,B)$ to denote the infinite-temperature OTOC averaged over all Hermitian operators $O_A$ (and $O_B$) supported in region $A$ (and $B$).
\eq{\OTOC(A,B)\equiv\mathop{\mathrm{avg}}\limits_{O_A,O_B}\frac{1}{D}\Tr O_A(t) O_B O_A(t) O_B.}
As shown in Ref.~\onlinecite{hosur2016chaos}, $\OTOC(A,B)$ can be expressed in terms of the 2nd R\'enyi entanglement features $W^{(2)}_{U(t)}[\sigma,\tau]$ following \eqnref{eq: def OTOC}. We will focus on the OTOC averaged over the unitary ensemble $\scE(t)$, which amounts to replacing $W^{(2)}_{U(t)}[\sigma,\tau]$ by its ensemble expectation $W^{(2)}[\sigma,\tau]$ given in \eqnref{eq: W2 large D} to the leading order of $D=d^N$. The choices of $\sigma$ and $\tau$ are specified in \eqnref{eq: st=AB}, which corresponds to $\overline{\sigma}=1-2N_A/N$, $\overline{\tau}=-1+2N_B/N$, $\overline{\sigma\tau}=-1+2(N_{A\cup B}-N_{A\cap B})/N$. Here $N_A$ (or $N_B$) is the size (number of qudits) of the operator support $A$ (or $B$). $N_{A\cap B}$ characterizes the operator overlap, and $N_{A\cup B}=N_A+N_B-N_{A\cap B}$.

With these, we found that to the leading order in $D$, the operator-averaged OTOC is given by
\eq{\label{eq: OTOC large D}\begin{split}
\OTOC&(A,B)= \scR_{[11\bar{1}\bar{1}]}d^{-2N_{A\cap B}}\\
&-2(\scR_{[11\bar{1}\bar{1}]}-\scR_{[2\bar{1}\bar{1}]})d^{-2N_{A\cup B}}\\
&+(\scR_{[00]}-\scR_{[11\bar{1}\bar{1}]})(d^{-2N_{A}}+d^{-2N_{B}})\\
&-(2\scR_{[00]}-\scR_{[0]}+2\scR_{[2\bar{1}\bar{1}]}-3\scR_{[11\bar{1}\bar{1}]})\\
&\hspace{12pt}\times(d^{-2(N_{A}+N_{B})}+d^{-2N})+\cdots.
\end{split}}
As expected, the expression is symmetric in exchanging $A$ and $B$. In the early time limit $t\to0$,
\eqs{\OTOC&=\frac{1}{d^{2N_{A\cap B}}}-2\kappa t^2+\scO(t^4),\\
\kappa&=\frac{1}{d^{2N_{A\cap B}}}+\frac{1}{d^{2N_{A\cup B}}}-\frac{1}{d^{2N_{A}}}-\frac{1}{d^{2N_{B}}}.}
The OTOC generally deviates quadratically ($\sim -2\kappa t^2$) from initial value in the early time, with $\kappa\geq 0$ for any choice of subsets $A$ and $B$ (except when $\kappa$ vanishes and the early time behavior will be taken over by $\scO(t^4)$ terms since OTOC has to be an even function of $t$ in our setting). The scrambling time is always of order 1 regardless of the operator size and the system size, due to the non-locality (not even k-local\cite{kitaev2002classical}) of the random Hamiltonian. In the late time limit $t\to\infty$, the OTOC approaches to the saturation value $\OTOC_\infty$ with oscillation. The envelop of the OTOC decays in power laws as
\eqs{&\OTOC\simeq\OTOC_\infty+\alpha t^{-9/2}+\beta t^{-6},\\
&\OTOC_\infty=\frac{1}{d^{2N_{A}}}+\frac{1}{d^{2N_{B}}}-\frac{1}{d^{2(N_{A}+N_{B})}}-\frac{1}{d^{2N}},\\
&\alpha=\frac{1}{\sqrt{2}\pi^{3/2}}\Big(\frac{1}{d^{2N_{A\cup B}}}-\frac{1}{d^{2(N_{A}+N_{B})}}-\frac{1}{d^{2N}}\Big),\\
&\beta=\frac{1}{\pi^2}\Big(\frac{1}{d^{2N_{A\cap B}}}+\frac{3}{d^{2(N_{A}+N_{B})}}-\frac{1}{d^{2N_{A}}}-\frac{1}{d^{2N_{B}}}\\
&\hspace{40pt}-\frac{2}{d^{2N_{A\cup B}}}+\frac{3}{d^{2N}}\Big).}
In all parameter regimes, it turns out that the $\alpha t^{-9/2}$ term is always overwhelmed by either the $\beta t^{-6}$ term  (for $t\lesssim t_d$) or the $\OTOC_\infty$ term (for $t\gtrsim t_d$), so the $t^{-9/2}$ will not be observed. Therefore the time scale that OTOC saturates to the final value will be set by the dip time $t_d=(\beta/\OTOC_\infty)^{1/6}$.

\begin{figure}[t]
\begin{center}
\includegraphics[width=0.92\linewidth]{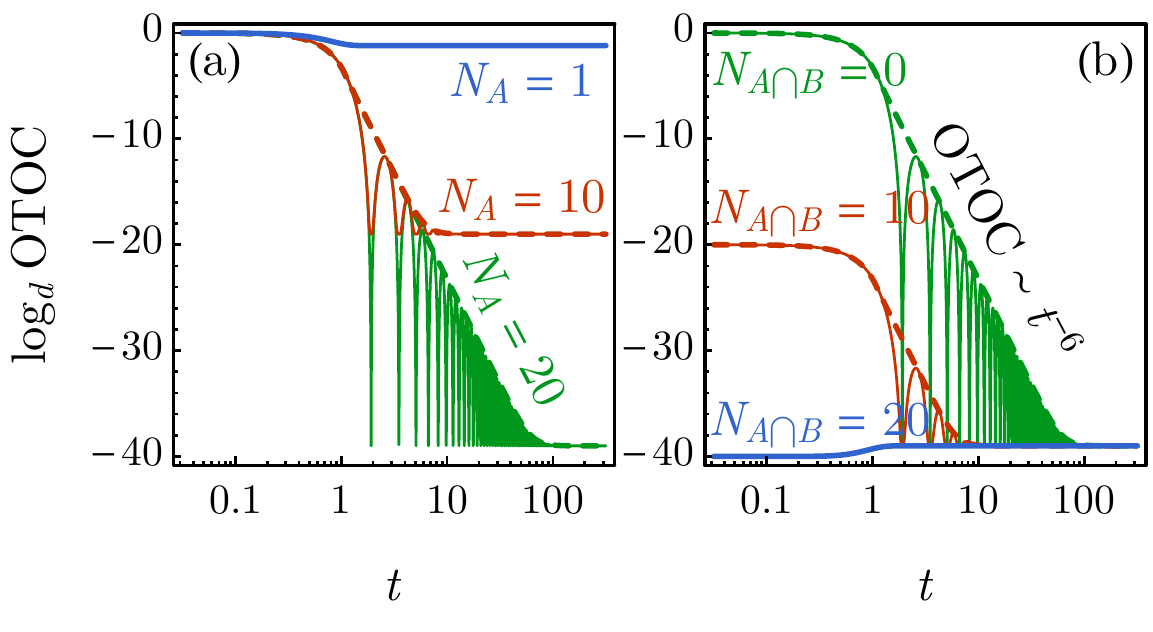}
\caption{Operator-averaged OTOC as a function of time in the logarithmic scale, for qubits ($d=2$) in the limit of $N\to\infty$. (a) The disjoint case $N_{A\cap B}=0$ with different operator size $N_A=N_B$. (b) Fixed operator size at $N_A=N_B=20$ with different overlap $N_{A\cap B}$. The solid curve is the OTOC and the dashed curves traces out the envelope function.}
\label{fig: OTOC}
\end{center}
\end{figure}

To be concrete, let us consider a more specific case when the operators $O_{A}$ and $O_{B}$ are of the same size $N_A=N_B$, and their supports overlap over $N_{A\cap B}$ qudits. In the limit that $d^{N_A}\ll d^N$, the typical behaviors of the operator-averaged OTOC are shown in \figref{fig: OTOC}. In the disjoint case \figref{fig: OTOC}(a) when $N_{A\cap B}=0$, the OTOC deviates from 1 as
\eq{\OTOC\simeq1-2 t^2+\cdots\quad(t\to 0),}
and approaches to saturation (to the leading order of $d$) as
\eq{\OTOC\simeq\frac{2}{d^{2N_A}}+\frac{1}{\pi^2 t^6}.}
The saturation time is $t_d=(d^{N_A}/\pi)^{1/3}/\sqrt{2}$. Within the intermediate time range $1\lesssim t\lesssim t_d$, the envelope of the OTOC exhibits the $t^{-6}$ power law behavior, which has been discussed in Ref.~\onlinecite{vijay2018finite,mertens2017solving, bagrets2017power}. Increasing the operator size $N_A$ will both suppress the saturation value and delay the saturation time exponentially, as shown in \figref{fig: OTOC}(a). Now if we fix the operator size and allows the operators $O_A$ and $O_B$ to overlap in their supports, the initial OTOC $d^{-2N_{A\cap B}}$ will be suppressed with $N_{A\cap B}$ exponentially, but the $t^{-6}$ power law behavior in the intermediate time range remains, as demonstrated in \figref{fig: OTOC}(b).

\subsection{Entanglement Growth after a Quench}

The entanglement features of the Hamiltonian generated unitary evolution can be applied to study the entanglement growth after a quantum quench.\cite{kim2013ballistic,liu2014entanglement,kaufman2016quantum} The quantum quench problem we will discuss here is to start with an initial \emph{product state} $\ket{\psi(0)}$ and evolve it by $U(t)=e^{-\ii H t}$ to the final state $\ket{\psi(t)}=U(t)\ket{\psi(0)}$. Generally, the quantum entanglement will grow in time and saturates to the thermal limit if the Hamiltonian is quantum chaotic. The entanglement features of the final state $\ket{\psi(t)}$ is all encoded in the entanglement features of $U(t)$. To reveal their relation, let us first define the entanglement features for a generic quantum many-body state $\ket{\psi}$ as\cite{you2018machine}
\eq{V_{\psi}^{(n)}[\tau]=\Tr (\ket{\psi}\bra{\psi})^{\otimes n} X_\tau,}
where $n$ is the R\'enyi index and $X_\tau$ is the representation of $\tau\in S_n^{\times N}$ in the $n$-replicated Hilbert space $\scH^{\otimes n}$. The entanglement features of a state is directly related to its entanglement entropies by
\eq{\label{eq: S=lnV}S_{\psi}^{(n)}[\tau]=\frac{1}{1-n}\ln V_{\psi}^{(n)}[\tau],}
with the entanglement region $A$ specified by the permutation $\tau$ following \eqnref{eq: A to cyc}.

If the state $\ket{\psi(t)}=U(t)\ket{\psi(0)}$ is obtained from the unitary evolution $U(t)$, the entanglement features of the state $V_{\psi(t)}^{(n)}[\tau]$ will be related to the entanglement features of the unitary evolution $W_{U(t)}^{(n)}[\sigma,\tau]$ by the following generic form
\eq{\label{eq: V=WPhi}V_{\psi(t)}^{(n)}[\tau]=\sum_{[\sigma]}W_{U(t)}^{(n)}[\sigma^{-1},\tau]\Phi^{(n)}_{\psi(0)}[\sigma],}
where $\Phi^{(n)}_{\psi(0)}[\sigma]$ is some function of $\sigma\in S_n^{\times N}$ that is determined by the initial state $\ket{\psi(0)}$. It is not the entanglement feature of the initial state, but can be determined from that, via
\eqs{\label{eq: V0=W0Phi} V_{\psi(0)}^{(n)}[\tau]&=\sum_{[\sigma]}W_{U(0)}^{(n)}[\sigma^{-1},\tau]\Phi^{(n)}_{\psi(0)}[\sigma]\\
&=\sum_{[\sigma]}d^{\#(\sigma^{-1}\tau)}\Phi^{(n)}_{\psi(0)}[\sigma],}
which is just an application of \eqnref{eq: V=WPhi} to $t=0$. Here $\#(g)$ denotes the cycle number of the permutation $g$. If the initial state is a \emph{product state}, we say that it is \emph{entanglement featureless}, in the sense that its entanglement features  $V_{\psi(0)}^{(n)}[\tau]=1$ are trivial constants, because the entanglement entropies of a product state always vanish for any choices of the permutation $\tau$ and the R\'enyi index $n$. Then from \eqnref{eq: V0=W0Phi}, we can find the solution of $\Phi_{\psi(0)}^{(n)}[\sigma]$,
\eq{\Phi_{\psi(0)}^{(n)}[\sigma]=\prod_{i}\sum_{\tau_i\in S_n}\Wg_{\tau_i^{-1}\sigma_i}=\frac{\Gamma(d)^N}{\Gamma(d+n)^N},}
where the Weingarten function $\Wg$ here is of the bound dimension $d$. Substitute the result back to \eqnref{eq: V=WPhi}, we obtain the relation between the entanglement features of $\ket{\psi(t)}$ and of $U(t)$,
\eq{\label{eq: V=W} V_{\psi(t)}^{(n)}[\tau]=\frac{\Gamma(d)^N}{\Gamma(d+n)^N}\sum_{[\sigma]}W_{U(t)}^{(n)}[\sigma^{-1},\tau].}
Therefore, the knowledge of the entanglement features of the unitary evolution itself is sufficient to determine how the entanglement will grow after a quantum quench from a product state, as we proposed in \eqnref{eq: Spsi}. A similar relation is also proposed in Ref.\,\onlinecite{Lensky2018} recently.

In the following, we will focus on the ensemble averaged 2nd-R\'enyi entanglement features of the state, 
\eq{V^{(2)}[\tau]=\langle V_{\psi(t)}^{(n)}[\tau]\rangle_{U(t)\in\scE(t)}.}
Applying \eqnref{eq: V=W} to \eqnref{eq: W2 large D}, we can obtain the entanglement features of $\ket{\psi(t)}$ to the leading order of $D=d^N$,
\eqs{\label{eq: V2 large D}V^{(2)}[\tau]&=V_\text{early}[\tau]+V_\text{late}[\tau],\\
V_\text{early}[\tau]&=\scR_{[11\bar{1}\bar{1}]}+\cdots,\\
V_\text{late}[\tau]&=\sum_{\upsilon=\pm1}D^{\frac{\upsilon\overline{\tau}-1}{2}}(\scR_{[00]}-\scR_{[11\bar{1}\bar{1}]})+\cdots,
}
where $\upsilon$ is an auxiliary Ising variable in the holographic bulk. As time evolves, $V^{(2)}[\tau]$ crosses over from the early-time behavior $V_\text{early}$ to the late-time behavior $V_\text{late}$. In the late time, the entanglement feature can be modeled by an Ising Hamiltonian $H_\text{late}$,
\eq{H_\text{late}[\tau;\upsilon]=-\frac{\ln d}{2}\sum_{i}\upsilon\tau_i,}
such that $V_\text{late}[\tau]=\sum_{[\upsilon]}e^{-H_\text{late}[\tau;\upsilon]}D^{-1/2}(\scR_{[00]}-\scR_{[11\bar{1}\bar{1}]})$. The holographic Ising model describes an holographic bulk variable $\upsilon$ couples to all qudit variables $\tau_i$. In terms of the random tensor network, this implies that the late-time state $\ket{\psi(t)}$ can be described by a big random tensor, which is consistent with the random matrix theory. 
From the perspective of  tensor network holography\cite{
evenbly2011tensor,
swingle2012entanglement,
qi2013exact}
, $\upsilon$ can be view as a black hole horizon in the sense that the Ryu-Takanayagi surface\cite{ryu2006holographic} can never cut through the interior of the random tensor that corresponds to $\upsilon$. Such a black hole picture naturally gives rise to the volume law entanglement entropy in the late time. 

\begin{figure}[t]
\begin{center}
\includegraphics[width=0.9\linewidth]{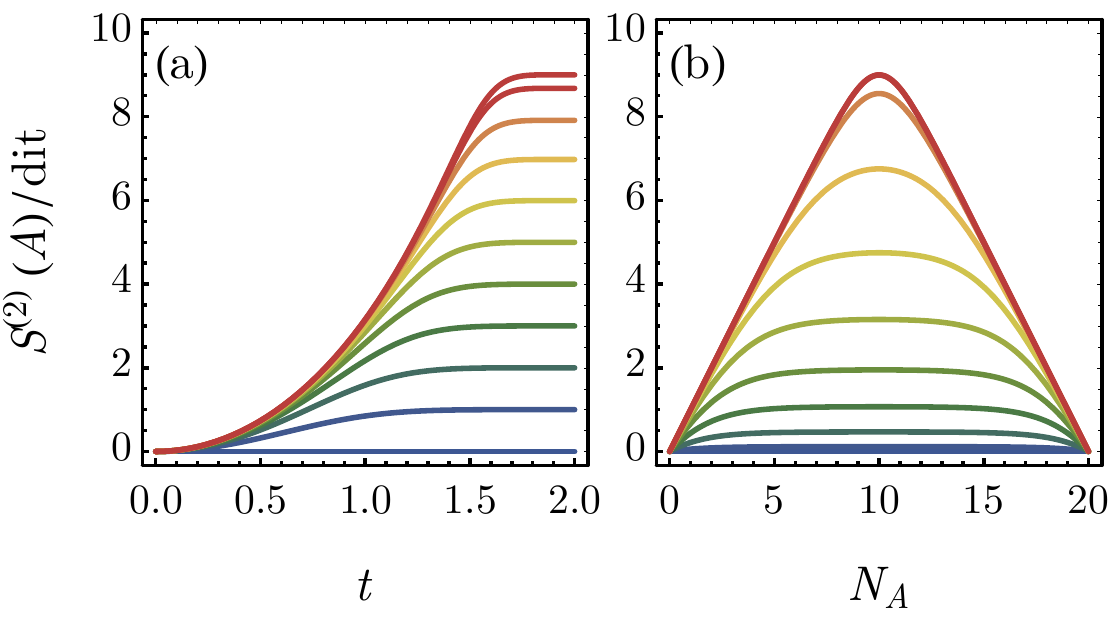}
\caption{A system of $N=20$ qudits. (a) The entanglement entropy grow after a quench from the product state for different size of the subset $A$, ranging from $N_A=0$ (blue) to $N_A=10$ (red). (b) The scaling of entanglement entropy with $N_A$ at different time, ranging from $t=0$ (blue) to $t=2$ (red).}
\label{fig: S grow}
\end{center}
\end{figure}

We can translate the entanglement feature to the entanglement entropy according to \eqnref{eq: S=lnV}. Given the result in \eqnref{eq: V2 large D}, the 2nd R\'enyi entropy over a subset $A$ of $N_A$ qudits of a quantum many-body state after a quench from the product state follows (to the leading $D$ order)
\eq{S^{(2)}(A)=-\ln(\scR_{[11\bar{1}\bar{1}]}+(1-\scR_{[11\bar{1}\bar{1}]})(d^{-N_A}+d^{-N_{\bar{A}}})),}
where $N_{\bar{A}}=N-N_A$. For a $N=20$ qudit system, we plot the entanglement entropy $S^{(2)}(A)$ as a function of both time $t$ and the subset size $N_A$ in \figref{fig: S grow}. 
The entropy grows quadratically in time as
\eq{S^{(2)}(A)=2(1-d^{-N_A}-d^{-N_{\bar{A}}})t^2+\scO(t^4),}
as shown in \figref{fig: S grow}(a).
In contrast to the linear growth of entropy for chaotic local Hamiltonian dynamics, the quadratic growth is a consequence of the non-locality of the random Hamiltonian we considered. In the late time, the entropy approaches to the ``volume law'' scaling,
\eq{S^{(2)}(A)\simeq N_A \ln d\quad(N_A\ll N/2),}
as shown in \figref{fig: S grow}(b).

\section{Summary}

In this work, we introduce the general concept of entanglement features that can be defined both for unitary gates $U$ as $W_{U}^{(n)}[\sigma,\tau]$ and for many-body states $\ket{\psi}$ as  $V_{\psi}^{(n)}[\tau]$, which provide a systematic characterization of their entanglement properties. 
In the simplest case (when $\sigma,\tau$ are cyclic), the entanglement features are just exponentiated entanglement entropies $\sim e^{(1-n)S^{(n)}}$. 
If we consider the entanglement entropy as a kind of ``free energy'' associated with the entanglement region, then the entanglement features are just the corresponding Boltzmann weights. From this perspective, the entanglement features describes a statistical ensemble of entanglement regions which encodes the ``features of entanglement'' in either a unitary gate or a many-body state. For more general permutations $\sigma,\tau\in S_n^{\times N}$, the entanglement features gives more refined description of quantum entanglement that can go beyond the description power of entanglement entropies. 

The entanglement feature is useful in relating many different ideas together. Many quantum information theoretic descriptions of entanglement, such as mutual and multi-partite informations, are unified within the framework of entanglement features. Moreover, several measures of quantum chaos including the out-of-time-order correlation and the entropy growth after quantum quench are all related to the entanglement features of the unitary evolution itself. 

At first glance, specifying the entanglement feature for every configuration of $\sigma$ and $\tau$ seems to involve exponentially large amount of data. However, the entanglement features are not independent. The hidden correlations among entanglement features allow a more efficient modeling with much less parameters. This idea is in parallel to compressing big data by neural network in machine learning where hidden neurons and deep neural networks are introduced to efficiently model the internal correlation among the data. Indeed in our study, the hidden Ising variables also naturally arise to simplify the model of entanglement features. These hidden variables can either be interpreted as hidden neurons in the neuron network language, or as the holographic bulk degrees of freedom in tensor network holography. The holographic picture provides us an intuitive understanding of the entanglement dynamics in quantum many-body systems.

To illustrate these general ideas, we take the unitary gates generated by random Hamiltonians as our example and calculate their ensemble-averaged entanglement features. For the 2nd-R\'enyi case, we are able to obtain analytic expressions for the entanglement features, from which we can find the underlying Ising model in the holographic bulk. This provides us a toy model to see the emergence of the holographic black hole under the random Hamiltonian dynamics. Finally we apply our results to study the OTOC and the entanglement growth under the random Hamiltonian dynamics. 
As a future direction, it will be interesting to generalize our approach to local Hamiltonians.

\begin{acknowledgements}
We acknowledge 
Tarun Grover,
Muxin Han,
Chao-Ming Jian,
Gil Refael
%Xiao-Liang Qi,
and Beni Yoshida
for helpful discussions. We thank Sagar Vijay and
Ashvin Vishwanath 
for helpful 
discussions and
sharing their draft  
Ref.~\onlinecite{vijay2018finite} before posting. 
We especially thank Xiao-Liang Qi for various enlightening discussions and valuable comments. 
YZY is supported by a Simons Investigator Grant of Ashvin Vishwanath. 
YG is supported by
the Gordon and Betty Moore Foundation EPiQS Initiative through Grant (GBMF-4306). 
\end{acknowledgements}

\appendix

\section{Entanglement Features of Two-Qudit Gates}
\label{appendix}

For two-qudit (i.e. $N=2$) gates we are able to investigate some aspects of entanglement features beyond ensemble average. In particular we found an interesting tentative bound for unitary two-qubit (i.e. $N=2,~d=2$) gate. %in subsection~\ref{subsection: bound}.

\subsection{Two-Qudit Cyclic Entanglement Features}

We consider the system with only two qudits (i.e. $N=2$), and label the input and output channels of the unitary gate by $A$, $B$, $C$ and $D$ as in \eqnref{eq: UABCD}. We focus on a subset of entanglement features, called the \emph{cyclic entanglement features}, where $\sigma_i,\tau_i$ can only take either the identity $\id$ or the cyclic $\cyc$ permutations. The cyclic entanglement features are sufficient to captures the entanglement entropies over all channels of the unitary gate to all R\'enyi indices. For two-qudit unitary gates, it can be shown that there are only two independent and non-trivial cyclic entanglement features, which can be denoted as $W_U^{(n)}(AC)$ and $W_U^{(n)}(AD)$,
\eq{\label{eq: UABCD}\dia{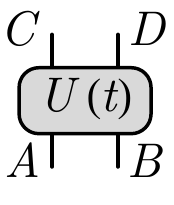}{42}{-19}\quad\begin{array}{c}
W_U^{(n)}(AC)\equiv W_U^{(n)}[\cyc\times\id,\cyc\times\id],\\
W_U^{(n)}(AD)\equiv W_U^{(n)}[\cyc\times\id,\id\times\cyc].
\end{array}}
The other cyclic entanglement features are either independent of $U$ (and thus trivial), such as
\eqs{W_U^{(n)}()&\equiv W_U^{(n)}[\id\times\id,\id\times\id]=d^{2n},\\
W_U^{(n)}(A)&\equiv W_U^{(n)}[\cyc\times\id,\id\times\id]=d^{n+1},\\
W_U^{(n)}(AB)&\equiv W_U^{(n)}[\cyc\times\cyc,\id\times\id]=d^{2};}
or are related the above mentioned entanglement features. Therefore, given the qudit dimension $d$, the cyclic entanglement features of two-qudit unitary gates can be fully characterized by two sets of entanglement entropies $S_U^{(n)}(AC)$ and $S_U^{(n)}(AD)$, which are directly related to the entanglement features $W_U^{(n)}(AC)$ and $W_U^{(n)}(AD)$ according to \eqnref{eq: S=lnW}.

\subsection{Entanglement Entropy Plot and Unitarity Bound}
\label{subsection: bound}

\begin{figure}[t]
\begin{center}
\includegraphics[width=0.9\linewidth]{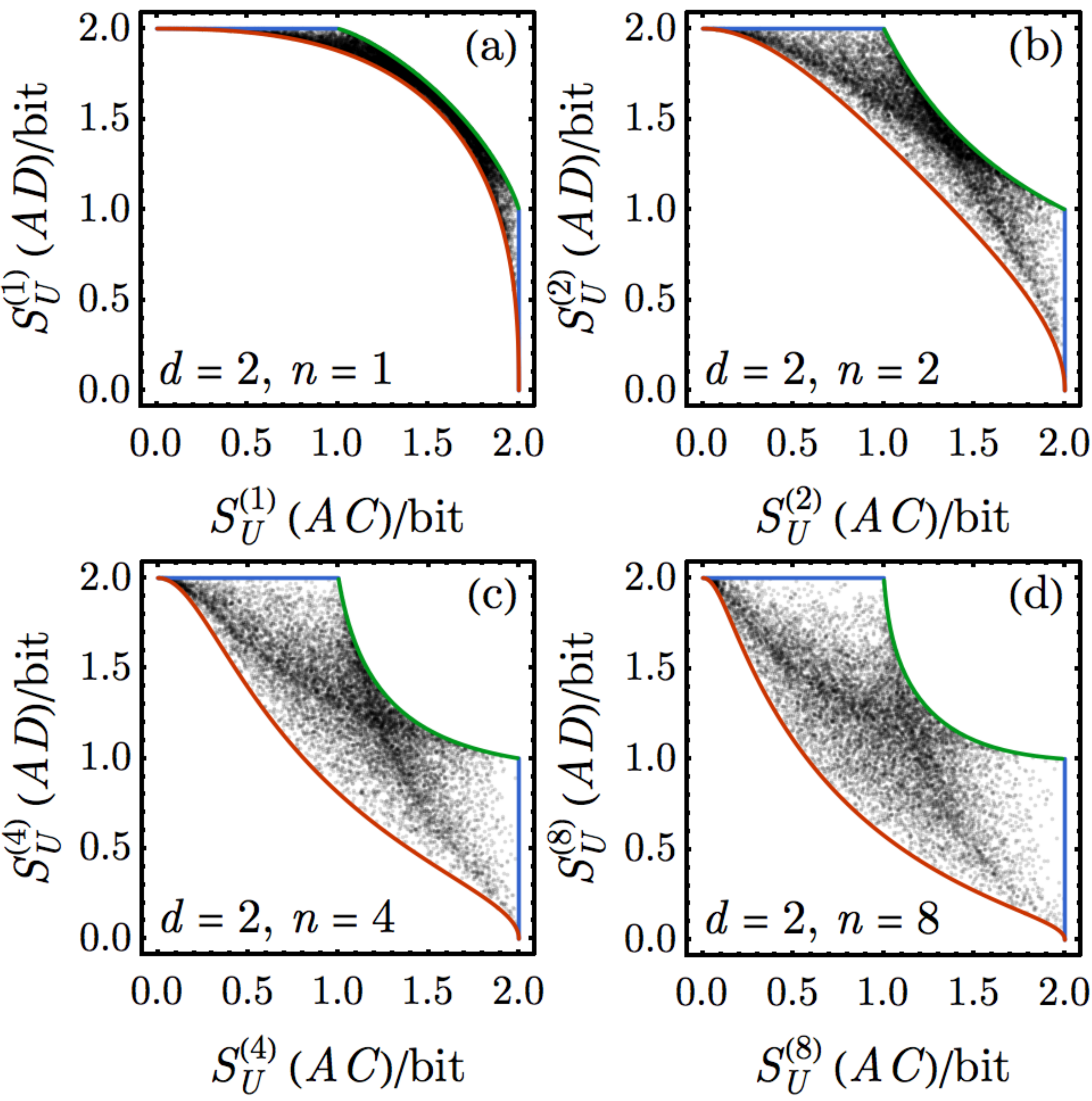}
\caption{Plot of entanglement entropies $S_U^{(n)}(AC)$ and $S_U^{(n)}(AD)$ for 10000 unitary gates $U$ generated by two-qubit ($N=2, d=2$) random Hamiltonians at late time. Cases of different R\'enyi indices $n$ are shown in different subfigures. The bounding curves are given by \eqnref{eq: Slow} (in red) and \eqnref{eq: Sup} (in green).}
\label{fig: Splot}
\end{center}
\end{figure}

We can numerically study the entanglement features of the unitary ensemble $\scE(t)$ generated by the random Hamiltonian. Each unitary $U$ drawn from the ensemble $\scE(t)$ can be represented as a point on the $S_U^{(n)}(AC)$-$S_U^{(n)}(AD)$ plane according to its entanglement features. Let us first consider the late-time unitary ensemble $\scE(t\to\infty)$ of two qubits (i.e. $N=2, d=2$), generated by
\eq{U=e^{-\ii H t},\quad H=\sum_{a,b}J_{ab}\sigma^{ab},}
where $a,b=0,1,2,3$ are Pauli indices and $\sigma^{ab}=\sigma^a\otimes\sigma^b$ form a basis of all two-qubit Hermitian operators. The coupling strength $J_{ab}$ are independently drawn from Gaussian distributions. The variance of $J_{ab}$ can be absorbed into $t$ as an (inverse) time scale, which is actually unimportant as we consider the late-time limit $t\to\infty$. These two-qubit unitaries are distributed in a bounded region of $S_U^{(n)}(AC)$ and $S_U^{(n)}(AD)$, as shown in \figref{fig: Splot}. The shape of the region changes with the R\'enyi index $n$. The entropies are measured in unit of dit $\equiv \ln d$ (which reduces to bit $\equiv \ln 2$ in the present case). 

\begin{table}[htbp]
\caption{Examples of the two-qubit Hamiltonian generated unitary $U=e^{-\ii H}$ at the conner points in the entropy region.}
\begin{center}
\begin{tabular}{ccc}
$H$ & $S_U^{(n)}(AC)$/dit & $S_U^{(n)}(AD)$/dit\\
\hline
$0$ & $0$ & $2$\\
$\frac{\pi}{4}\sigma^{11}$ & $1$ & $2$ \\
$\frac{\pi}{4}(\sigma^{11}+\sigma^{22})$ & $2$ & $1$ \\
$\frac{\pi}{4}(\sigma^{11}+\sigma^{22}+\sigma^{33})$ & $2$ & $0$ \\
\end{tabular}
\end{center}
\label{tab: H}
\end{table}

These entropy regions are pinned to four corner points at $(S_U^{(n)}(AC),S_U^{(n)}(AD))=(0,2), (1,2), (2,1), (2,0)$ bit. Representative Hamiltonians that generate such unitaries at these corner points are provided in \tabref{tab: H}. It turns out that the boundary of the entropy region can be traced out by connecting these corner points by linearly interpolating the representative Hamiltonians. For example, in terms of the entanglement features, the lower edge of the region (red curve in \figref{fig: Splot}) is given by
\eqs{\label{eq: Slow}W_U^{(n)}(AC)&=8^{-n}(3 (1-\cos t)^n+(5+3\cos t)^n),\\
W_U^{(n)}(AD)&=8^{-n}(3 (1+\cos t)^n+(5-3\cos t)^n),}
for $t\in[0,\pi/4]$, and the upper edge of the region (green curve in \figref{fig: Splot})  is given by
\eqs{\label{eq: Sup}W_U^{(n)}(AC)&=2^{1-n}(\cos^{2n}t+\sin^{2n}t),\\
W_U^{(n)}(AD)&=2^{1-2n}((1-\sin2t)^n+(1+\sin2t)^n),}
for $t\in[0,\pi/4]$. The side edges connecting $(0,2)$ to $(1,2)$ and $(2,0)$ to $(2,1)$ are simply straight lines (blue segments in \figref{fig: Splot}). The upper and lower edges are non-linear bounds on entanglement entropies for unitary gates, which clearly goes beyond the previously known linear bounds, such as the subaddtivitiy and strong subadditivity relations.

We need to remark here that the tentative unitarity bounds for R\'enyi entropies are observations based on numerics. It will be interesting to find a solid proof for these bounds on the sub-region R\'enyi entropies of two-qubit unitary gates. 
In addition, our method of finding these boundary curves by interpolating special Hamiltonians does not generalized to $d>2$ cases. The corresponding non-linear bound for general two-qudit ($d>2$) unitary gates remains an open question. 

\subsection{Large-$d$ Limit and Entropy Trajectory}

\begin{figure}[t]
\begin{center}
\includegraphics[width=0.9\linewidth]{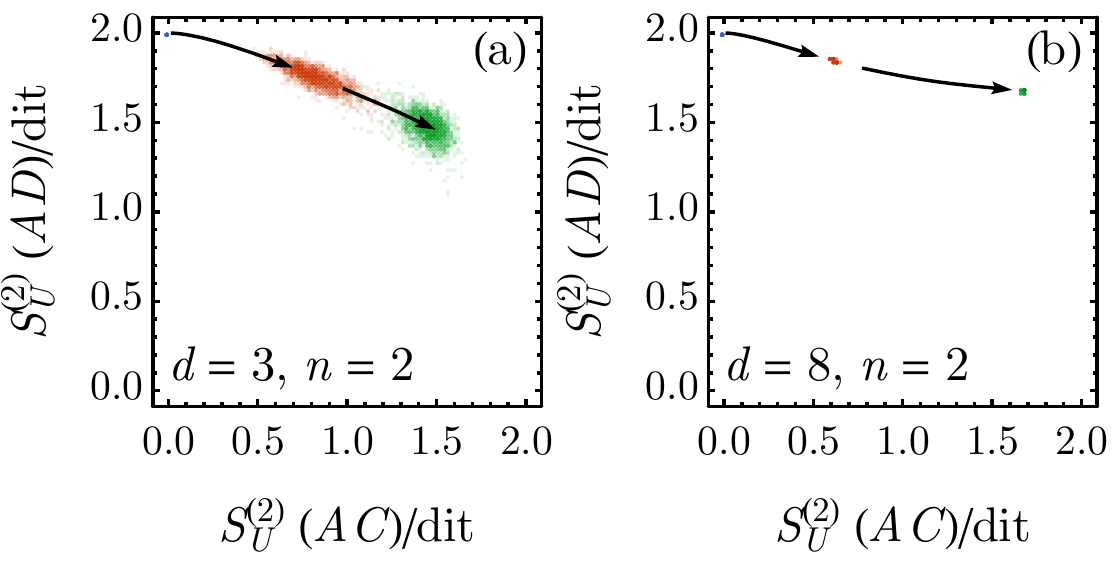}
\caption{Time evolution of the unitary ensemble $\scE(t)$ on the entropy plane of $S_U^{(n)}(AC)$ and $S_U^{(n)}(AD)$ for two-qudit system of (a) $d=3$ and (b) $d=8$. The blue, red and green points are respectively the initial ($t=0$), intermediate ($t=0.8$) and late-time ($t\to\infty$) ensembles.}
\label{fig: trajectory}
\end{center}
\end{figure}

Similar numerical study of the entanglement features under the two-qudit random Hamiltonian dynamics can be carried out for $d>2$ cases. \figref{fig: trajectory} shows the time evolution of the ensemble $\scE(t)$ on the entropy plane of $S_U^{(n)}(AC)$ and $S_U^{(n)}(AD)$ (we only shows the results of $n=2$, for other R\'enyi indices have very similar behaviors). Each point in the plot corresponds to a unitary gate $U$ drawn from the ensemble $\scE(t)$.

We can see that the fluctuation of the entanglement entropies $S_U^{(n)}$ is quickly suppressed as the qudit dimension $d$ gets larger. For $d=3$, the fluctuation of ensemble $\scE(t\rightarrow \infty)$ is already too small to efficiently map out the unitarity bound in numerics. 
For $d=8$ in \figref{fig: trajectory}(b), the fluctuation is negligible that the ensemble basically traces out a well-defined entropy trajectory under the random Hamiltonian dynamics. This implies that we can study the dynamics of the ensemble averaged entanglement features $\langle W_U^{(n)}\rangle_{U\in\scE(t)}$ in the large $d$ limit.

\subsection{Averaged Entanglement Features of Two Qudits}

\begin{figure}[t]
\begin{center}
\includegraphics[width=\linewidth]{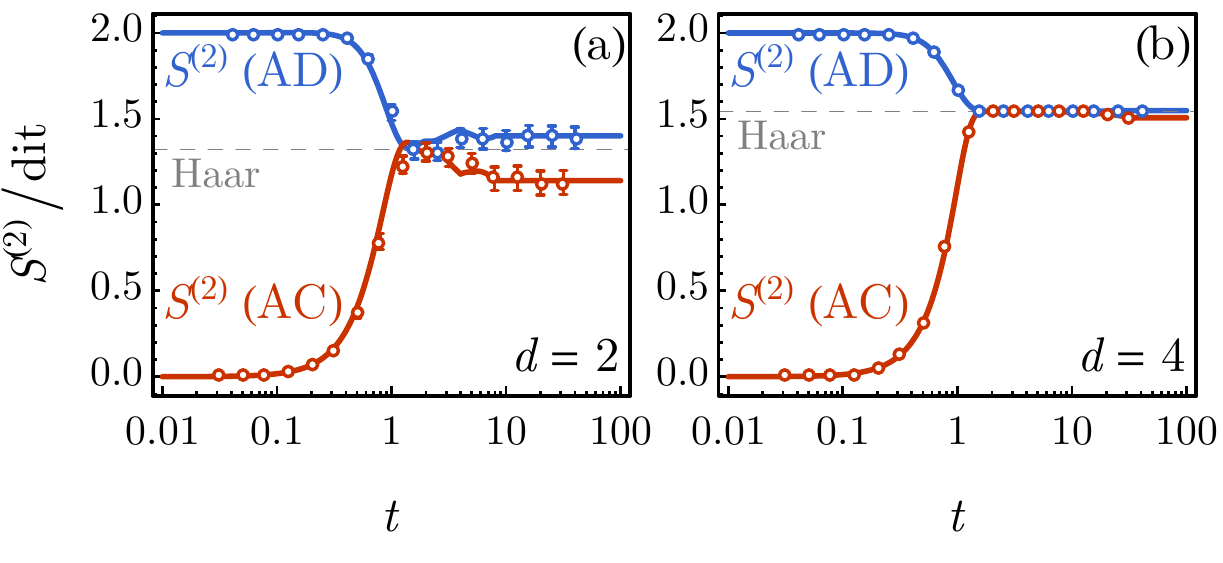}
\caption{Ensemble averaged 2nd-R\'enyi entropies $S^{(2)}$ of two-qudit unitaries for (a) $d=2$ and (b) $d=4$. The small circles are numerical simulations on 1000 random Hamiltonians. The curves are theoretical results based on \eqnref{eq: W2 result} (without taking the large $D$ limit). The $S^{(2)}$ for Haar random unitaries are marked out by dashed lines according to \eqnref{eq: W2 Haar2}.}
\label{fig: S curve}
\end{center}
\end{figure}

According to \eqnref{eq: W2 result}, we can analytically calculate the two non-trivial 2nd-R\'enyi entanglement features $W^{(2)}(AC)$ and $W^{(2)}(AD)$ for the two-qudit random Hamiltonian dynamics, as defined in \eqnref{eq: UABCD}. The result is presented in terms of the Renyi entropies $S^{(2)}(AC)$ and $S^{(2)}(AD)$ and benchmarked with numerics in \figref{fig: S curve}. The theoretical curves matches the numerical data points nicely.
The late-time limit of the entanglement features can be calculated from \eqnref{eq: W2 late},
\eqs{W_\infty^{(2)}(AC)&=2\Big(d^2+\frac{1}{d^2+1}-\frac{4}{d^2+3}\Big),\\
W_\infty^{(2)}(AD)&=2\Big(d^2-1+\frac{3}{d^2+1}-\frac{4}{d^2+3}\Big),}
both of which deviate from that of Haar unitaries given by \eqnref{eq: W2 Haar}
\eq{\label{eq: W2 Haar2}W_\text{Haar}^{(2)}(AC)=W_\text{Haar}^{(2)}(AD)=\frac{2d^4}{d^2+1}.}
The amount of deviation reduces with the qudit dimension $d$ and becomes negligible in the large $d$ limit, as can be seen by comparing \figref{fig: S curve}(a) and (b). In conclusion, the entropy trajectory under the random Hamiltonian dynamics in two-qudit system can be traced out based on \eqnref{eq: W2 result}. 

\section{Exact Result of Entanglement Features}\label{sec:exact}

\subsection{Weingarten Functions}

The exact form of the Weingarten function on $S_n$ group is given by (for $D\geq n$)
\eq{\Wg_g=\frac{1}{(n!)^2}\sum_{\lambda}\frac{\chi^{\lambda}(1)^2\chi^{\lambda}(g)}{s_{\lambda,D}(1)},}
where the sum is over all partitions $\lambda$ of $n$. Here $\chi^\lambda$ is the character of $S_n$ corresponding to the partition $\lambda$ and $s_{\lambda,D}$ is the Schur polynomial of $\lambda$ such that $s_{\lambda,D}(1)$ is simply the dimension of the representation of $U(D)$ corresponding to $\lambda$. The Weingarten function is a class function, which means it only depends on the cycle type $\nu(g)$.

For $S_4$ group, the Weingarten functions can be enumerated as in \tabref{tab: Wg S4}. They have a common denominator, which will be denoted as
\eq{\label{eq: Z4}Z_4(D)=\prod_{k=0}^4(D^2-k^2).}
Using this result, we can carry out the $S_4$ group summation in \eqnref{eq: W2 result} exactly and obtain the ensemble averaged 2nd R\'enyi entanglement feature $W^{(2)}$ to all orders of $D$.
\begin{table}[htbp]
\caption{Weingarten functions on $S_4$ group.}
\begin{center}
\begin{tabular}{cc}
$\nu(g)$ & $\Wg_g$\\
(1,1,1,1) & $(D^4-8D^2+6)/Z_4(D)$\\
(2,1,1) & $-D(D^2-4)/Z_4(D)$\\
(3,1) & $(2D^2-3)/Z_4(D)$\\
(4) & $-5D/Z_4(D)$\\
(2,2) & $(D^2+6)/Z_4(D)$ 
\end{tabular}
\end{center}
\label{tab: Wg S4}
\end{table}

\subsection{Entanglement Features (Exact)}
To all orders of $D$, the 2nd R\'enyi entanglement feature still take the early-time and late-time form as in \eqnref{eq: W=W+W} and \eqnref{eq: W=DF}, which we repeat here
\eq{\begin{split}
&W^{(2)}[\sigma,\tau]=W_\text{early}[\sigma,\tau]+W_\text{late}[\sigma,\tau],\\
&W_\text{early}[\sigma,\tau]=\sum_{\upsilon=\pm1}D^{\frac{1}{2}(\upsilon\overline{\sigma\tau}+\upsilon)}F_\text{early}(\upsilon),\\
&W_\text{late}[\sigma,\tau]=\sum_{\upsilon_{1,2}=\pm1}D^{\frac{1}{2}(\upsilon_1\overline{\sigma}+\upsilon_2\overline{\tau}+\upsilon_1\upsilon_2)}F_\text{late}(\upsilon_1\upsilon_2).
\end{split}}
The common denominator $Z_4(D)$ of the Weingarten function defined in \eqnref{eq: Z4} can be factored out, which allows us to define
\eq{F_\text{early}(\upsilon)=\frac{f_\text{early}(\upsilon)}{Z_4(D)},\quad F_\text{late}(\upsilon)=\frac{f_\text{late}(\upsilon)}{Z_4(D)}.}
Now we present the exact result for $f_\text{early}(\upsilon)$ and $f_\text{late}(\upsilon)$ as follows
\begin{widetext}
\eq{\begin{split}
f_\text{early}(+1)=&D^3\big(
4(D^2+6)(\scR_{[00]}-\scR_{[0]})+16(2D^2-3)\scR_{[1\bar{1}]}+(D^2-3)(D^2-4)\scR_{[2\bar{2}]}-4D^2(D^2+1)\scR_{[1\bar{1}0]}\\
&-4D^2(D^2-4)\scR_{[2\bar{1}\bar{1}]}+D^2(D^2-3)(D^2-4)\scR_{[11\bar{1}\bar{1}]}\big),\\
f_\text{early}(-1)=&2D^5\big(
10(\scR_{[0]}-\scR_{[00]})-4(D^2+1)\scR_{[1\bar{1}]}-(D^2-4)\scR_{[2\bar{2}]}\\
&+4(2D^2-3)\scR_{[1\bar{1}0]}+(D^2-3)(D^2-4)\scR_{[2\bar{1}\bar{1}]}-D^2(D^2-4)\scR_{[11\bar{1}\bar{1}]}\big),\\
f_\text{late}(+1)=&D^{9/2}\big(
-2(D^2-14)\scR_{[0]}+(D^4-11D^2+8)\scR_{[00]}-40\scR_{[1\bar{1}]}-(D^2-4)\scR_{[2\bar{2}]}+4(D^2+6)\scR_{[1\bar{1}0]}\\
&+6(D^2-4)\scR_{[2\bar{1}\bar{1}]}-D^2(D^2-4)\scR_{[11\bar{1}\bar{1}]}\big),\\
f_\text{late}(-1)=&D^{9/2}\big(
(D^2+1)(D^2-12)\scR_{[0]}-2(D^4-12D^2+12)\scR_{[00]}+8(D^2+6)\scR_{[1\bar{1}]}+3(D^2-4)\scR_{[2\bar{2}]}\\
&-20D^2\scR_{[1\bar{1}0]}-2D^2(D^2-4)\scR_{[2\bar{1}\bar{1}]}+3D^2(D^2-4)\scR_{[11\bar{1}\bar{1}]}\big).\end{split}}
The spectral form factors $\scR_{[k]}$ were calculated in Ref.~\onlinecite{cotler2017chaos}. We copy it here for the completeness of the presentation.
\eq{\begin{split}
\scR_{[0]}(t)=&\scR_{[00]}(t)=1,\\
\scR_{[1\bar{1}]}(t)=&\scR_{[1\bar{1}0]}(t)=r_1(t)^2+(1-r_2(t))/D,\\
\scR_{[2\bar{2}]}(t)=&\scR_{[1\bar{1}]}(2t),\\
\scR_{[21\bar{1}]}=&r_1(2t)r_1(t)^2+(-r_1(2t)r_2(t)r_3(2t)-2r_1(t)r_2(2t)r_3(t)++r_1(2t)^2+2r_1(t)^2)/D\\
&+(2 r_2(3t)-r_2(2t)-2r_2(t)+1)/D^2,\\
\scR_{[11\bar{1}\bar{1}]}=&r_1(t)^4+(-2 r_1(t)^2
r_2(t)r_3(2t)-4r_1(t)^2r_2(t)+2r_1(2t)r_1(t)^2+4r_1(t)^2)/D\\
&+(2r_2(t)^2+r_2(t)^2r_3(2t)^2+8r_1(t)r_2(t)r_3(t)-2r_1(2t)r_2(t)r_3(2t)-4r_1(t)r_2(2t)r_3(t)\\
&\hspace{12pt}+r_1(2t)^2-4r_1(t)^2-4r_2(t)+2)/D^2\\
&+(-7r_2(2t)+4r_2(3t)+4r_2(t)-1)/D^3,
\end{split}}
where the functions $r_{1,2,3}(t)$ are defined as
\eq{
r_1(t)=\frac{J_1(2t)}{t},\quad 
r_2(t)=\Big(1-\frac{|t|}{2D}\Big)\Theta\Big(1-\frac{|t|}{2D}\Big),\quad 
r_3(t)=\frac{\sin(\pi t/2)}{\pi t/2}.}
\end{widetext}

\bibliography{ref.bib}
\end{document}